\documentclass[aps,prd,twocolumn,
               notitlepage,mathrsfs,
               amssymb,amsmath,amsfonts,
               superscriptaddress,
               longbibliography,
               nofootinbib,floatfix]{revtex4-2}

\usepackage{graphicx}
\usepackage{url}
\usepackage[linktocpage,breaklinks]{hyperref}
\usepackage[capitalize]{cleveref}
\usepackage[usenames,dvipsnames]{xcolor}
\hypersetup{colorlinks=true,
            citecolor=NavyBlue,
            linkcolor=magenta,
            urlcolor=magenta}
\usepackage{multirow,array}
\DeclareMathAlphabet{\pazocal}{OMS}{zplm}{m}{n}
\usepackage{journals}
\newcommand{\tb}[1]{\left( \hat{#1} \right)}
\newcommand{\pa}{\partial}
\newcommand{\tg}{\tilde \gamma}
\newcommand{\tG}{\tilde \Gamma}
\newcommand{\tA}{\tilde A}

\begin{document}

\title{SACRA-2D: New axisymmetric general relativistic hydrodynamics code with fixed mesh refinement}

\date{\today}

\author{Alan Tsz-Lok Lam}
\email{tszlok.lam@aei.mpg.de}
\affiliation{Max Planck Institute for Gravitational Physics (Albert Einstein Institute), 14476 Potsdam, Germany}

\author{Masaru Shibata}
\affiliation{Max Planck Institute for Gravitational Physics (Albert Einstein Institute), 14476 Potsdam, Germany}
\affiliation{Center of Gravitational Physics and Quantum Information, Yukawa Institute for Theoretical Physics, Kyoto University, Kyoto, 606-8502, Japan} 

\begin{abstract}
We present \texttt{SACRA-2D}, a new MPI and OpenMP parallelized, fully relativistic hydrodynamics (GRHD) code in dynamical spacetime under axial symmetry with the cartoon method using the finite-volume shock-capturing schemes for hydrodynamics. Specifically, we implemented the state-of-the-art HLLC Riemann solver and found better accuracy than the standard Total Variation Diminishing Lax-Friedrich Riemann solver.
The spacetime evolves under the Baumgarte-Shapiro-Shibata-Nakamura formalism with Z4c constraint propagation. We demonstrate the accuracy of the code with some benchmark tests and excellent agreement with other codes in the literature. A wide variety of test simulations, including the head-on collision of black holes, the migration and collapse of neutron stars, and the collapse of a rotating supermassive star to a massive black hole and a disk, is also performed to show the robustness of our new code. 
\end{abstract}

\maketitle

\section{Introduction}

In hydrodynamics and magnetohydrodynamics simulations, the finite volume method with the high-resolution shock-capturing (HRSC) scheme is commonly used due to its conservative nature and capability to resolve sharp discontinuities, such as shocks, that often appear in the fluid's motion.
One popular HRSC scheme is the family of the Harten, Lax and van Leer (HLL) based approximate Riemann solver \cite{hart83},
which utilizes a subset of waves in the Riemann fan.
While most existing numerical relativity (magneto)hydrodynamics codes (e.g., \cite{most14,etie15,bern16,most19,viga20,cipo21,fouc21,haya22,radi22})
employ the HLLE solver \cite{kurg00} that includes only shocks and rarefactions,
it is known to be very diffusive \cite{toro94,mign05,mign09,held18} and the accuracy for long-term simulation could be deteriorated.
This is relevant for modeling the long-term evolution of post-merger remnant from neutron-star mergers \cite{gao25}, particularly important when considering the magnetohydrodynamical processes~\cite{kiuc22}.
The authors in \cite{whit16,kiuc22} have reported a new implementation of the HLLC solver,
which is a more sophisticated Riemann solver that restores the contact discontinuity in the Riemann fan.
A recent study also demonstrates its significance 
even in the inspiral phase of the binary neutron stars, 
where the dynamical tidal effect on the gravitational waveform can only be manifested with the HLLC solver \cite{kuan24}.
Consequently, employing the HLLC solver (or a more accurate solver) for astrophysical simulations is crucial for accurate (magneto)hydrodynamics and gravitational wave signals.

Despite many relativistic astrophysical systems requiring spatially three-dimensional simulation to fully capture the dynamics,
such numerical studies are usually computationally expensive, prohibiting us from studying a wider range of parameters.
On the other hand, we could approximate specific systems to be axisymmetric, reducing the problem's size to two spatial dimensions and drastically lowering the computation cost for numerical simulation.
This allows us to follow the physical system for a much longer time scale beyond the current capability of three-dimensional simulations.
Indeed, axisymmetric GRHD code with dynamical spacetime has been used extensively to study various astrophysical systems,
such as the dynamics of isolated neutron stars \cite{kiuc08,shib03b,shib03c} and hypermassive neutron stars \cite{duez04, Duez:2005sf, Duez:2005cj, Shibata:2005mz, Duez:2006qe, shib17},
stellar collapse \cite{seki04,seki05,shib04a,shib06,kuro12}
and collapsar scenario \cite{fuji21,fuji23a,fuji24,shib24,shib25},
collapse of supermassive stars \cite{Shibata:2002br, Liu2007oct, mont12,shib16b,uchi17,uchi19a,shib16b,fuji24b},
black hole-torus system \cite{mont08,mont10,shib12},
higher-dimensional spacetime \cite{yosh09}, and 
merger remnants from binary neutron stars and
black hole-neutron star \cite{fuji17,fuji20a,fuji20b,shib21,shib21b,fuji23b}.

In addition to astrophysical events, there has recently been an increasing interest in numerical relativity simulations of modified theories of gravity,
aiming to search for distinctive features in the strong field regime that may provide shreds of evidence with current and future observations.
In particular, a major effort has been put into analyzing the properties of compact objects, including black holes and neutron stars,
as well as investigating the gravitational wave signals from the coalescence of binary compact objects in modified gravity theories such as the scalar-tensor theory (STT) \cite{shib14,tani15,kuan23b,lam24a,lam24b},
the scalar Gauss-Bonnet theory \cite{wite19,ripl19,ripl20,silv21,east22,kuan23a},
the dynamical Chern-Simon gravity \cite{dels15,okou19,okou20,rich23},
and the STT with kinetic screening \cite{beza22,shib23}.
However, the three-dimensional setups are computationally too costly to perform numerical experiments to survey new theories systematically,
which is particularly important in exploring a well-posed formulation for certain theories.
While one-dimensional simulation has been vastly used to explore the effect of the modification in gravity (e.g., \cite{mend16,sper17,cheo19,kuan21a,kuan21b,kuan22}), 
axisymmetric GRHD code can act as a bedrock for an efficient alternative to implementing various alternative theories of gravity
and helping to gain new intuition in the regime of non-zero angular momentum.
The cartoon method has been shown to be very useful for studying long-term dynamics, for example,
core-collapse supernova in STT \cite{kuro23} and the superradiant instability of a Proca field \cite{east17a,east17b}.

This paper reports the implementation \texttt{SACRA-2D}, a new MPI and OpenMP parallelized, fully relativistic GRHD code in dynamical spacetime under axial symmetry with the cartoon method. 
The code is written in FORTRAN90 with the numerical algorithm and technique closely resembling the three-dimensional moving box numerical relativity code \texttt{SACRA-MPI} \cite{yama08,kiuc17}.
We implemented the Baumgarte-Shapiro-Shibata-Nakamura (BSSN) formalism \cite{shib95,baum98} with Z4c constraint propagation \cite{bern10,hild13} to solve Einstein's  equations. 
The finite-volume shock-capturing scheme is employed for GRHD.
Specifically, we implemented the TVDLF solver and the state-of-the-art HLLC solver for the approximate Riemann solver.

In the following, we first outline the grid structure of \texttt{SACRA-2D} in \cref{sec:grid}.
We then describe the implementation for dynamical spacetime in \cref{sec:bssn},
specifically the details of the cartoon method in \cref{sec:cartoon},
followed by the formulation for GRHD in \cref{sec:GRHD}.
In \cref{sec:result}, we validate our code with several benchmark test problems,
addressing the accuracy and performance of \texttt{SACRA-2D}.
The parallelization efficiency is discussed in \cref{sec:scaling}.
Unless specified otherwise, we adopt the geometric unit of $G=c=1$ throughout this paper, where $G$ and $c$ are the gravitational constant and speed of light, respectively. 
The subscripts $a$, $b$, $c, \cdots$ denote the spacetime coordinates while $i$, $j$, $k, \cdots$ the spatial coordinates, respectively. 

\section{Formulation}

\subsection{Grid structure} \label{sec:grid}
The grid setting of \texttt{SACRA-2D} is very similar to that of the "box-in-box" simulation \cite{brug08,yama08}.
We employ the two-to-one fixed mesh refinement (FMR) structure in the computational domain, which is composed of a hierarchy of nested concentric grids overlaying on top of each other.
It consists of $L$ levels of FMR domains,
each of which contains an even number of grids $N$ in both $x$ and $z$ directions with the grid spacing written as
\begin{align}
    \Delta x^{(0)} &= x_{\max} / N, &
        \Delta z^{(0)} &= z_{\max} / N, \nonumber \\
    \Delta x^{(l)} &= \Delta x^{(l-1)} / 2, &
    \Delta z^{(l)} &= \Delta z^{(l-1)} / 2, 
\end{align}
for $l=1,2,\cdots,L-1$, where $x_{\max}$ and $z_{\max}$ are the size of computational domain, and levels $0$ and $(L-1)$ represent the coarsest and  finest levels, respectively.
The metric and hydrodynamics variables are assigned at cell-centered positions with coordinates
\begin{align}
    x_j^{(l)} &= \left( j - \frac{1}{2} \right) \Delta x^{(l)}, &
    z_k^{(l)} &= \left( k - \frac{1}{2} \right) \Delta z^{(l)},
\end{align}
for $j,k \in [1,N]$ on the $l$-th FMR level.
The cell interfaces $x_{j \pm 1/2}^{(l)}$ and $z_{k\pm 1/2}^{(l)}$ are located at
 $x_j^{(l)} \pm \Delta x^{(l)}/2$ and  $z_k^{(l)} \pm \Delta z^{(l)}/2$, respectively.
 
In addition to the local $N$ grid cells, extra buffer cells are necessary for calculating derivatives with finite different schemes and reconstructing the hydrodynamics variables.
For sixth-order accuracy in spatial derivative, four buffer zones are required to handle the lopsided finite difference for the advection term (see \cref{sec:bssn}) as well as the prolongation scheme at the refinement boundary.
We also allocate an additional four buffer cells on top of the original four buffer zones to facilitate the adaptive time update in the time integration scheme (see \cref{sec:refinement} for more details).
Therefore, in \texttt{SACRA-2D}, we set up a total of $(4+4)$ buffer cells in each direction for the purpose of time integration.
However, the number of buffer cells can be easily adjusted if a higher/lower order scheme is used [e.g., $(3+3)$ for fourth-order accuracy].

\subsection{Einstein's equations}\label{sec:bssn}
\subsubsection{Basic equations}
Einstein's equations are first formulated under the standard Arnowitt–Deser–Misner (ADM) $(3+1)$ formulation \cite{arno62,york79}, in which the line element is written in the form of
\begin{align}
    ds^2 &= - \alpha^2 dt^2 + \gamma_{ij}
    \left( dx^i + \beta^i dt \right)
    \left( dx^j + \beta^j dt \right),
\end{align}
where $\alpha$ and $\beta^i$ are the lapse function and shift vector, respectively,
and spatial three-metric $\gamma_{ij}$ is defined from the spacetime metric $g_{ab}$ as
\begin{align}
    \gamma_{ab} := g_{ab} + n_a n_b,
\end{align}
with $n_a$ being a time-like unit normal vector orthogonal to the space-like hypersurface.

Then, following the Baumgarte-Shapiro-Shibata-Nakamura (BSSN) formalism \cite{shib95,baum98} with Z4c constraint propagation \cite{bern10,hild13},
we reformulate the field equations defining the following set of geometric variables in Cartesian coordinates, 
\begin{subequations}
\begin{align}
    \tg_{ij} &:= \psi^{-4} \gamma_{ij}, &
    h_{ij} &:= \tg_{ij} - f_{ij}, &
    W &:= \psi^{-2}, 
\end{align}
\begin{align}
    K &:= \gamma_{ij} K^{ij}, &
    \tA_{ij} &:= \psi^{-4} \left( K_{ij} - \frac{1}{3} \gamma_{ij} K \right), \\
    \tG^i &:= - \pa_j \tg^{ij}, &
    \hat K &:= K - 2 \Theta,
\end{align}
\end{subequations}
where $\psi$ is the conformal factor,
$\tg_{ij}$ is the conformal spatial metric,
$h_{ij}$ is the residual of spatial metric,
$f_{ij}$ is the time-independent flat background metric,
$K_{ij}$ is the extrinsic curvature,
$\Theta:= - n_a Z^a$ is a constraint in Z4 system \cite{bona03,bona04,gund05}, and
$\hat K$ is a variable used for the evolution equations.

The evolution equations for the geometric variables in Cartesian coordinates are given by
\begin{subequations} \label{eq:bssn}
\begin{align}
    ( \pa_t - &\beta^k \pa_k ) W =
        \frac{1}{3} W \left[ \alpha \left(\hat K + 2 \Theta \right) - \pa_k \beta^k \right], \\
\begin{split}
    ( \pa_t - &\beta^k \pa_k ) h_{ij} = 
        - 2 \alpha \tA_{ij} \\
        &+ \tg_{ik} \pa_j \beta^k
        + \tg_{jk} \pa_i \beta^k - \frac{2}{3} \tg_{ij} \pa_k \beta^k,
\end{split}
\end{align}
\begin{align}
\begin{split}
    ( \pa_t - &\beta^k \pa_k ) \tA_{ij} = 
        W^2 \left[ \alpha R_{ij} -  D_i D_j \alpha - 8\pi \alpha S_{ij} \right]^{\rm TF} \\
        &+ \alpha \left[ \left(\hat K + 2 \Theta \right) \tA_{ij} - 2 \tA_{ik} \tA_j{}^k \right] \\
        &+ \tA_{kj} \pa_i \beta^k + \tA_{ki} \pa_j \beta^k
        - \frac{2}{3} \tA_{ij} \pa_k \beta^k,
\end{split} \\
\begin{split} \label{eq:K} 
    (\pa_t - &\beta^k \pa_k) \hat K
        = 4\pi \alpha (S^i{}_i+E) 
        + \alpha \kappa \Theta \\
        &+ \alpha \left[ \tA_{ij} \tA^{ij} + \frac{1}{3} \left(\hat K + 2 \Theta \right)^2 \right]
        -D_i D^i \alpha, 
\end{split}
\end{align}
\begin{align}
\begin{split}
    (\pa_t - &\beta^k \pa_k ) \tG^i =
        - 2 \tA^{ij} \pa_j \alpha
        + 2 \alpha \left[ \tG^{i}_{jk} \tA^{jk} \right. \\
        &\left. - \frac{1}{3} \tg^{ij} \pa_j \left(\hat 2 K + \Theta \right)
        - \frac{3}{W} \tilde A^{ij} \pa_j W
        - 8 \pi \tg^{ij} S_j \right] \\
        &+ \frac{2}{3} \tilde \gamma^{jk} \tilde\Gamma^i{}_{jk} \pa_l \beta^l 
        + \tg^{jk} \pa_j \pa_k \beta^i 
        + \frac{1}{3} \tg^{ij} \pa_j \pa_k \beta^k \\
        &- \tilde \gamma^{kl} \tilde \Gamma^j{}_{kl} \pa_j \beta^i
        - 2 \alpha \kappa \left( \tG^i - \tilde \gamma^{kl} \tilde \Gamma^j{}_{kl} \right), 
\end{split} \\
\begin{split}
    (\pa_t - &\beta^k \pa_k ) \Theta =
        \frac{1}{2} \alpha \left[ 
        R - \tA_{ij} \tA^{ij}
        + \frac{2}{3} \left(\hat K + 2 \Theta \right)^2
        \right] \\
        &- 8 \alpha \pi E - 2 \alpha \kappa \Theta ,
\end{split}
\end{align}
\end{subequations}
where $D_i$ and $R_{ij}$ are the covariant derivative and the Ricci tensor associated with $\gamma_{ij}$, respectively,
${\rm TF}$ corresponds to the trace-free part of the tensor,
and $(E, S_i, S_{ij})$ are the $3+1$ decomposition of 
the stress-energy tensor $T_{ab}$ defined by
\begin{subequations}
\begin{align}
    E &:= n^a n^b T_{ab}, \\
    S_i &:= - \gamma_i{}^a n^b T_{ab}, \\
    S_{ij} &:= \gamma_i{}^a \gamma_j{}^b T_{ab}.
\end{align}
\end{subequations}
The constraint damping parameter $\kappa$ is chosen to be $\kappa = 5 \times 10^{-3}M^{-1}$ in this work with $M$ being the total mass of the system.
We also enforce the following algebraic constraints during the evolution
\begin{equation}
    \det (\tg_{ij})= 1~~~\mathrm{and}~~~
    \tg^{ij} \tA_{ij}= 0,
\end{equation}
as the numerical error could induce violation in these constraints.
Specifically, we reset the metric variables after each time integration given by
\begin{subequations}
\begin{align}
    \tg_{ij}^{\rm new} &= \det (\tg_{ij})^{-1/3} \tg_{ij}, \\
    W^{\rm new} &= \det (\tg_{ij})^{-1/6} W, \\
    \tA_{ij}^{\rm new} &= \det (\tg_{ij})^{-1/3} \left( 
        \tA_{ij} - \frac{1}{3} \tg_{ij} \tg^{kl} \tA_{kl} \right), \\
    K^{\rm new} &= K + \tg^{ij} \tA_{ij},
\end{align}
\end{subequations}
to satisfy the algebraic constraints.

We adopt the standard moving-puncture gauge condition \cite{alcu03,bake06,camp06} for the lapse function and shift vector as
\begin{subequations} \label{eq:moving_puncture}
\begin{align}
    ( \pa_t - \beta^j \partial_j ) \alpha &= - 2 \alpha K , \\
    ( \pa_t - \beta^j \partial_j ) \beta^i &= \frac{3}{4} B^i , \\
    ( \pa_t - \beta^j \partial_j ) B^i &= ( \pa_t - \beta^j
    \pa_j ) \tilde{\Gamma}^i - \eta_B B^i,
\end{align}
\end{subequations}
where $B^i$ is an auxiliary variable and $\eta_B$ is a constant damping factor typically chosen to be $\simeq 1/M$ with $M$ being the total mass of the system.
We employ the standard initial gauge choice for the lapse function $\alpha = \psi^{-2}$ and the shift vector $\beta^i = 0 = B^i$ for all the tests in \cref{sec:result} unless stated otherwise.

The spatial derivatives in the right-hand side of BSSN equations, \cref{eq:bssn}, are evaluated with a sixth-order central finite difference,
while the sixth-order lopsided finite difference is used for the advection terms in the left-hand side of \cref{eq:bssn} to guarantee the stability.
To reduce high-frequency noise, we include eighth-order Kreiss-Oliger (KO) dissipation for geometric variables $Q$ in $x$ and $z$ directions as
\begin{align}
    \left( \varepsilon / 256 \right) \left(\Delta x^8 \pa_x^8 + \Delta z^8 \pa_z^8\right) Q,
\end{align}
with the damping parameter $\varepsilon$ set to be $0.5$.

\subsubsection{Cartoon method}\label{sec:cartoon}
We employ the cartoon method \cite{shib00,alcu01, shib03} to impose the axial symmetry on the geometric variables defined in the Cartesian coordinates.
Three extra layers of the computational domain are constructed upon and below the $x$-$z$ plane with $y = \pm j \Delta y$ ($j=1, \cdots, 3$) as required by the sixth-order central finite difference.
Einstein's equations are solved only on the $y=0$ plane while the geometric variables $Q(x, \pm j \Delta y, z)$ on the $y = \pm j \Delta y$ planes are obtained by first interpolating the variables $Q^{(0)}(\varpi, 0, z)$ 
at the same radial distance $\varpi:= \sqrt{x^2 + \left( j\Delta y \right)^2}$ on the $y=0$ plane
using Lagrange’s formula with nine nearby points $[x_j-4\Delta x, x_j+4\Delta x]$ along the $x$ direction 
and then apply rotation using the assumption of axial symmetric as
\begin{equation}
\begin{aligned}
    Q &= Q^{(0)}, &
    Q_{z} &= Q_{z}^{(0)}, \\
    Q_{A} &= \Lambda_{A}{}^{B} Q_{B}^{(0)}, &
    Q_{zz} &= Q_{zz}^{(0)}, \\
    Q_{Az} &= \Lambda_{A}{}^{B} Q_{Bz}^{(0)}, &
    Q_{AB} &= \Lambda_{A}{}^{C} \Lambda_{B}{}^{D} Q_{CD}^{(0)}, 
\end{aligned}
\end{equation}
where $Q$, $Q_i$, and $Q_{ij}$ denote, respectively, the scalar, vector, and tensor types of geometric variables, in BSSN formulation,
and $\Lambda_{A}{}^B$ is the rotational matrix given by
\begin{align}
    \Lambda_{A}{}^B &= 
    \begin{pmatrix}
        \cos \phi & - \sin\phi \\
        \sin \phi & \cos \phi
    \end{pmatrix},
\end{align}
with $\tan \phi := \pm j \Delta y / \varpi$. Note that the subscripts $A$ and $B$ run $x$ or $y$.
The interpolated values with eighth-order accuracy result in an expected sixth-order accuracy in the second derivative,
and allow us to compute the spatial derivative in the $y$-direction using the finite difference scheme in the same manner as in 3D Cartesian coordinates.
In particular, we enforce the derivatives $\{\pa_y, \pa_{xy}, \pa_{yz}\}$ on $\{Q, Q_z, Q_{zz}\}$ to be zero
in all equations to avoid double precision errors arising from the arithmetic operation of finite difference.
As we set $\Delta y^{(l)} = \Delta x^{(l)}$ for all FMR levels, the interpolation coefficients remain the same across all the FMR levels.
Hence, the coefficients can be easily pre-computed and saved for later use to speed up the calculation.

Since the neighboring nine points $x_j \pm 4\Delta x$ are required for the interpolation,
the geometric variables located on the extra layers $y=\pm j \Delta y$ at the edge of the FMR level with grid points $x \in [N+5, N+8]$ cannot be determined,
which causes trouble in obtaining the $xy$-derivative $\pa_{xy}$ for grid points $x \in [N+1, N+4]$.
To avoid this problem, we instead adopt the following form
\begin{subequations}\label{eq:cartoon_bnd}
\begin{align}
    \pa_{xy} Q_x &=   \frac{Q_y}{x^2} - \frac{\pa_x Q_y}{x}, &
    \pa_{xy} Q_y &= - \frac{Q_x}{x^2} + \frac{\pa_x Q_x}{x}, 
\end{align}
\begin{align}
\begin{split}
    \pa_{xy} Q_{xz} &=   \frac{Q_{yz}}{x^2} - \frac{\pa_x Q_{yz}}{x}, \\
    \pa_{xy} Q_{yz} &= - \frac{Q_{xz}}{x^2} + \frac{\pa_x Q_{xz}}{x}, \\
    \pa_{xy} Q_{xx} &=   2 \left( \frac{Q_{xy}}{x^2} - \frac{\pa_x Q_{xy}}{x} \right), \\
    \pa_{xy} Q_{yy} &= - 2 \left( \frac{Q_{xy}}{x^2} - \frac{\pa_x Q_{xy}}{x} \right), \\
    \pa_{xy} Q_{xy} &= \frac{Q_{yy} - Q_{xx}}{x^2} 
        + \frac{\pa_x Q_{xx} - \pa_x Q_{yy}}{x},
\end{split}
\end{align}
\end{subequations}
for the vector $Q_i$ and tensor $Q_{ij}$ quantities located at grid points $x \in [N+1,N+4]$.
Although the coordinate singularity $1/x$ appears in the source term of \cref{eq:cartoon_bnd},
it is justified since the grid points $x \in [N+1,N+4]$ are located at the edge of the refinement boundary far from the symmetric axis with non-zero $x$.

\subsubsection{Boundary condition}
For the outer boundary, we impose the outgoing boundary condition \cite{shib95} for metric variables $Q$ 
located at radial distance $r$ in the form
\begin{align}
    Q^n(r) = \left( 1 - \frac{\Delta t}{r} \right) Q^{n-1}( r - \Delta t),
\end{align}
in order to preserve $rQ$ along the characteristic curves $r-t = {\rm constant}$.
Here, $Q^n$ and $Q^{n-1}$ are variables in the current $t$ and previous $t-\Delta t$ time step, respectively, and we interpolate $Q^{n-1}$ at $ r - \Delta t$ with second-order Lagrange interpolation. \\

Since the Z4c prescription allows the propagation and damping of constraints by introducing the auxiliary variable $\Theta$,
constraint violation will be induced at the outer boundary and propagate inwards if the boundary condition above is used.
While one could avoid this by implementing constraint preserving boundary condition \cite{ruiz11},
we instead adopt a simple treatment for $\Theta$ following \cite{kyut14}.
We set an effective radius $r_{\rm Z4}$, beyond which the damping parameter $\kappa$ and the source term for $\Theta$ are multiplied by an additional factor  $\exp[- (x^2 + z^2) / r_{\rm Z4}^2]$ to suppress the propagation of constraint violation terms exponentially.
We typically set as $r_{\rm Z4} \lesssim L_{\rm max}/6$ equivalent to a factor of $\sim 10^{-16}$ at the outer boundary,
which corresponds to the same order of error as double precision.
We found that this simple treatment is good enough to maintain a stable evolution for the long term without any significant growth in constraint violation.

\subsection{General relativistic hydrodynamics} \label{sec:GRHD}

\subsubsection{Basic equations}

This section briefly summarizes the formulation for general relativistic hydrodynamics (GRHD) under 3+1 decomposition.
We refer readers to \cite{rezz13, shib16} for more detailed derivation.

The evolution equations for GRHD are based on the conservation of rest-mass and stress-energy momentum tensor,
\begin{subequations}
\begin{align}
    \nabla_a \left( \rho u^a \right) &= 0, \\
    \nabla_b T^{ab} &= 0,
\end{align}
\end{subequations}
where $\rho$, $u^a$, and $P$ are the rest-mass density, four-velocity, and pressure of the fluid, respectively,
and 
\begin{align}
    T_{ab} := \rho h u_a u_b + P g_{ab}
\end{align}
is the stress-energy tensor for perfect fluid
with $h:=1 + \epsilon + P/\rho$ being the specific enthalpy and $\epsilon$ being the specific internal energy. $\nabla_a$ denotes the covariant derivative with respect to $g_{ab}$.

We adopt the finite volume method using the formulation of, e.g., ~\cite{bany97} in the reference metric formalism \cite{mont14,cheo21} to solve the hydrodynamical system in cylindrical coordinates $(\varpi, \phi, z)$ at $\phi = 0$ plane.
Under such formulation, the GRHD equations can be written in the following conservative form
\begin{align} \label{eq:cons_form}
    \pa_t \mathbf{q} + \frac{1}{\sqrt{\hat\gamma}} \pa_i \left( \sqrt{\hat\gamma} \mathbf{f}^i \right) = \mathbf{s},
\end{align}
where $\hat \gamma_{ij}$ is the time-independent reference metric chosen to be flat metric in cylindrical coordinates; here $\hat \gamma_{ij}:= f_{ij}={\rm diag}(1, \varpi, 1 )$,
$\hat\gamma := \det(\hat \gamma_{ij})$ is the determinant,
$\mathbf{q} := \left( q_D, q_{S_i}, q_{E} \right)$ are the conservative variables defined as
\begin{align} \label{eq:cons_var}
\begin{pmatrix}
    q_D \\
    q_{S_i} \\
    q_{E}
\end{pmatrix}
= \psi^6
\begin{pmatrix}
    D \\
    {S_i} \\
    {E}
\end{pmatrix}
= \psi^6
\begin{pmatrix}
    \rho w \\
    \rho h w u_i \\
    \rho h w^2 - P
\end{pmatrix},
\end{align}
$\mathbf{f^i}$ are the flux terms written as
\begin{align}
\mathbf{f^i} &=
\begin{pmatrix}
    \left( f_D \right)^i\\
    \left( f_{S_j} \right)^i \\
    \left( f_{E} \right)^i
\end{pmatrix}
= \alpha \psi^6
\begin{pmatrix}
    D \bar v^i \\
    S_j \bar v^i + P \delta_j{}^i \\
    E \bar v^i + P \left( \bar v^i + \displaystyle\frac{\beta^i}{\alpha} \right)
\end{pmatrix},
\end{align}
with $w:= - n_a u^a = \alpha u^t$ being the Lorentz factor measure by an Eulerian observer
and $\bar v^i := -\beta^i+\gamma^{ij}u_j/u^t$. 

Since Einstein's equations are solved in Cartesian coordinates $(x, y, z)$ at $y=0$ plane,
the hydrodynamic variables can be rewritten in Cartesian coordinates as
$\varpi = x$ and $u_\phi = x u_y$,
which is essentially the same as the conversion to orthonormal frame in reference metric approach \cite{baum13}.
The source term $\mathbf{s}:= \left( s_D, s_{S_i}, s_{E} \right)$ in \cref{eq:cons_form} can then be evaluated in Cartesian coordinates in the forms
\begin{subequations}
\begin{align}
    s_D &= 0 \\
    \begin{split}
    s_{S_\varpi} &= P \pa_x \left( \alpha W^{-3} \right)
       - W^{-3} \rho h w^2 \left[ 
       \pa_x \alpha - \bar v_i \pa_x \beta^i \right.\\
       &\left. + \frac{W^2}{2} \alpha \bar v_i \bar v_j \pa_x \tg^{ij}
       + \frac{4}{W} \alpha \bar v_i \bar v^i \pa_x W
       \right]
       + \frac{(f_{S_\phi})^{\phi}}{\varpi} ,
    \end{split}\\
    s_{S_\phi} &= 0 \\
    \begin{split}
    s_{S_z} &= P \pa_z \left( \alpha W^{-3} \right)
       - W^{-3} \rho h w^2 \left[ 
       \pa_z \alpha - \bar v_i \pa_z \beta^i \right.\\
       &\left. + \frac{W^2}{2} \alpha \bar v_i \bar v_j \pa_z \tg^{ij}
       + \frac{4}{W} \alpha \bar v_i \bar v^i \pa_z W
       \right]
    \end{split}\\
    s_{E} &= \alpha K^{ij} S_{ij} - W^{-3} \rho h w^2 \bar v^i  \pa_i \alpha,
\end{align}
\end{subequations}
where the final term $\displaystyle\frac{(f_{S_\phi})^{\phi}}{\varpi}$ in $s_{S_\varpi}$ comes from the cylindrical geometry (see \cite{cheo21} for detailed derivation of geometrical source term).
Note that we have the conservation of angular momentum in axial symmetry.
In the conservative form of \cref{eq:cons_form} with $s_D=0=s_{S_\phi}$,
the conservation of mass and angular momentum can be satisfied numerically with machine precision.

Here, we write down the explicit discretized form of the volume-averaged equations in cylindrical coordinates as follows: 
\begin{align}
\begin{split}
	\pa_t &\left< \mathbf{q} \right>_{j,k} = \left< \mathbf{s} \right>_{j,k} - \frac{1}{\Delta V_{j,k}} \\
		&\times  \left\{ 
		\left[ \left< \mathbf{f} \right>^\varpi_{j+\frac{1}{2},k} \Delta A^\varpi_{j+\frac{1}{2},k} - 
		\left< \mathbf{f} \right>^\varpi_{j-\frac{1}{2},k} \Delta A^\varpi_{j-\frac{1}{2},k} \right] \right. \\
		&+   \left.
		\left[ \left< \mathbf{f} \right>^z_{j,k+\frac{1}{2}} \Delta A^z_{j,k+\frac{1}{2}} -
		\left< \mathbf{f} \right>^z_{j,k-\frac{1}{2}} \Delta A^z_{j,k-\frac{1}{2}} \right] \right\}
\end{split}
\end{align}
where $\Delta V_{j,k}$ and $\Delta A^i_{j,k}$ are the volume and the surface area of the cell $(j,k)$, respectively, given by
\begin{subequations}
\begin{align}
\begin{split}
	&\Delta V_{j,k} = 2\pi \int_{x_{j-\frac{1}{2}}}^{x_{j+\frac{1}{2}}} 
            \int_{z_{k-\frac{1}{2}}}^{z_{k+\frac{1}{2}}} x dx dz
		= 2 \pi x_j \Delta x \Delta z,
\end{split} \\
\begin{split}
	&\Delta A^\varpi_{j\pm \frac{1}{2},k} =
		2\pi \int_{z_{k-\frac{1}{2}}}^{z_{k+\frac{1}{2}}} x_{j\pm \frac{1}{2}} dz 
		= 2 \pi x_{j \pm \frac{1}{2}} \Delta z,
\end{split}\\
\begin{split}
	&\Delta A^z_{j,k\pm \frac{1}{2}} =
		2\pi \int_{x_{j-\frac{1}{2}}}^{x_{j+\frac{1}{2}}} x dx 
		= 2 \pi x_{j} \Delta x,
\end{split}
\end{align}
\end{subequations}
$\left< \mathbf{q} \right>_{j,k}$ and  $\left< \mathbf{s} \right>_{j,k}$ are the volume averaged of the corresponding quantities,
and $\left< \mathbf{f} \right>^i$  are the surface-averaged quantities of the flux terms at the cell interfaces.

\subsubsection{Riemann Solver}
We adopt the HSRC scheme to handle the flux term in hydrodynamics equations.
Both Total Variation Diminishing Lax-Friedrich (TVDLF) \cite{yee89,toth96,kepp12} and HLLC \cite{toro94,batt97,mign05} approximate Riemann solvers are implemented in \texttt{SACRA-2D}.
To obtain the numerical flux,
we first reconstruct the left and right states of the primitive variables $\mathbf{p}=(\rho, u_i, P, \epsilon)$ with 3rd-order piecewise parabolic method (PPM) \cite{cole84,shib05} at the cell interface.
Since the metric variables are smooth, we employ Lagrangian interpolation to calculate the values at the interface.
For the HLLC solver, we perform the tetrad transformation \cite{whit16,kiuc22} at the cell interface after reconstruction to obtain the numerical flux.

Here, we briefly outline the procedure of the HLLC solver
and refer readers to \cite{whit16,kiuc22,xie24} for more details on the implementation.
To evaluate the numerical flux in the $x$-direction,
we define a tetrad basis \cite{whit16,kiuc22} on the surface of $x_{j \pm 1/2}$ as
\begin{subequations}
\begin{align}
   & e_{\tb{t}}^a = n^a
       = \frac{1}{\alpha} \left( 1, - \beta^i \right), &&\\
   & e_{\tb{x}}^a = W \hat{B} \left( 0, \tg^{xi} \right),&& \\
   & e_{\tb{y}}^a = W \hat{D} \left( 0, 0, \tg_{zz}, - \tg_{yz} \right),&& \\
   & e_{\tb{z}}^a = W \hat{C} \left( 0, 0, 0, 1 \right),&&
\end{align}
\end{subequations}
where 
\begin{align}
   \hat B &= \frac{1}{\sqrt{\tg^{xx}}}, &
   \hat C &= \frac{1}{\sqrt{\tg_{zz}}}, &
   \hat D &= \frac{1}{\sqrt{\tg^{xx}\tg_{zz}}} = \hat B \hat C,
\end{align}
with the corresponding covariant components written as
\begin{subequations}
\begin{align}
   & e_{\tb{t}}{}_a = n_a = - \left( \alpha, 0, 0, 0 \right), \\
   & e_{\tb{x}}{}_a = W^{-1} \hat{B} 
       \left( \beta^x, 1, 0, 0 \right), \\
   & e_{\tb{y}}{}_a = W^{-1} \hat{D} 
       \left( - \tg_{xy} \beta^x + \tg_{xx} \beta^y, - \tg_{xy}, \tg_{xx}, 0 \right), \\
   & e_{\tb{z}}{}_a = W^{-1} \hat{C} \left( \beta_z, \tg_{iz} \right).
\end{align}
\end{subequations}
This allows us to transform the primitive variables from the Eulerian frame $\mathbf{p}$ to the tetrad frame $\mathbf{\tilde p}$ by
\begin{align}
	u_{\tb{a}} &= e_{\tb{a}}{}^b u_b, \\
	w^2 &= 1 + u^{\tb{i}} u_{\tb{i}}, \\
	v_{\tb{i}} &= u_{\tb{i}} / w.
\end{align}
We can then obtain the left (L) and right (R) states of the conservative variables $\mathbf{\tilde q}_{L/R}:=\mathbf{\tilde q} (\mathbf{\tilde p}_{L/R})$
and the flux terms $\mathbf{\tilde f}_{L/R}:=\mathbf{\tilde f}^{\tb{x}} (\mathbf{\tilde p}_{L/R})$ in the tetrad frame from the
corresponding left/right states of the primitive variables $\mathbf{\tilde p}_{L/R}$ as 
\begin{align}
\mathbf{\tilde q} (\mathbf{\tilde p}) &= 
\begin{pmatrix}
	D \\ 
	S_{\tb{j}} \\
	E
\end{pmatrix} 
=
\begin{pmatrix}
   \rho w \\
   \rho h w v_{\tb{j}} \\
   \rho h w^2 + P
\end{pmatrix}, \\
\mathbf{\tilde f}^{\tb{x}} (\mathbf{\tilde p}) &=
\begin{pmatrix}
   \left( \tilde f_{ D} \right)^{\tb{x}}\\
   \left( \tilde f_{ S_{\tb{j}}} \right)^{\tb{x}} \\
   \left( \tilde f_{ E} \right)^{\tb{x}}
\end{pmatrix} 
=
\begin{pmatrix}
   D v^{\tb{x}} \\
   S_{\tb{j}} v^{\tb{x}} + P \delta_{\tb{j}}^{\tb{x}} \\
   \left( E + P \right) v^{\tb{x}}
\end{pmatrix}, 
\end{align}
which essentially have the same expression as in special relativistic hydrodynamics.
Now, we can employ the HLLC solver in the local Minkowski spacetime \cite{mign05} to calculate the numerical flux as
\begin{align}
\mathbf{\tilde f}^{\tb{x}} = 
\begin{cases}
	\mathbf{\tilde f}^{\tb{x}}_L		~&\text{for } \lambda_L > v^{\tb{x}}_{\rm interface} \\
	\mathbf{\tilde f}^{\tb{x}}_{cL}	~&\text{for } \lambda_L < v^{\tb{x}}_{\rm interface} < \lambda_c\\
	\mathbf{\tilde f}^{\tb{x}}_{cR}	~&\text{for } \lambda_c < v^{\tb{x}}_{\rm interface} < \lambda_R\\
	\mathbf{\tilde f}^{\tb{x}}_R		~&\text{for } \lambda_R < v^{\tb{x}}_{\rm interface}
\end{cases},
\end{align}
where $\lambda_c$ is the characteristic speed of the contact discontinuity, 
$v^{\tb{x}}_{\rm interface} = {\beta^x} /\left( \alpha \sqrt{\gamma^{xx}}\right)$ is the interface velocity \cite{whit16,kiuc22},
$\mathbf{\tilde f}^{\tb{x}}_{cL/cR}$ and $\mathbf{\tilde q}^{\tb{x}}_{cL/cR}$ are the intermediate states obtained from the jump condition
\begin{align}
	\mathbf{\tilde f}^{\tb{x}}_{cL/cR} = \mathbf{\tilde f}^{\tb{x}}_{L/R}
		+ \lambda_{L/R} \left( \mathbf{\tilde q}^{\tb{x}}_{cL/cR} - \mathbf{\tilde q}^{\tb{x}}_{L/R} \right),
\end{align}
and $\lambda_{L/R}$ are the left/right characteristic speed given by
\begin{align}
	\lambda_L &= \min(\lambda(\mathbf{\tilde{p}}_L)^-, \lambda(\mathbf{\tilde{p}}_R)^-), \\
	\lambda_R &= \max(\lambda(\mathbf{\tilde{p}}_L)^+, \lambda(\mathbf{\tilde{p}}_R)^+), \\
\begin{split}
	\lambda^\pm (\mathbf{\tilde{p}}) &= \frac{1}{1-v^2 c_s^2}
		\left\{ v^{\tb{x}} \left( 1- c_s^2 \right) \right.\\
		&\left. \pm c_s \sqrt{ \left(1-v^2 \right) \left[
		1 - v^2 c_s^2 - \left(1 -c_s^2 \right) v^{\tb{x}2}
   		 \right]} \right\},
\end{split}
\end{align}
with $v^2 := v^{\tb{i}} v_{\tb{i}}$ and $c_s$ being the sound speed.
The characteristic speed $\lambda_c$ can be obtained by imposing the continuity condition of the pressure across the contact discontinuity as \cite{mign05}
\begin{align}
\left(\tilde f_{E}^{\rm HLL}\right)^{\tb{x}} \lambda_c^2
- \left[ E^{\rm HLL} + \left( \tilde f_{S_{\tb{x}}}^{\rm HLL} \right)^{\tb{x}} \right] \lambda_c
    + S_{\tb{x}}^{\rm HLL} = 0,
\end{align}
where $\mathbf{\tilde q}^{\rm HLL}$ and $\mathbf{\tilde f}^{\tb{x}, \rm HLL}$ represent the HLL state of the conserved quantities and flux, respectively, given by
\begin{align}
    \mathbf{\tilde q}^{\rm HLL} &= 
        \frac{\lambda_R \mathbf{\tilde q}_R - \lambda_L \mathbf{\tilde q}_L
        + \mathbf{\tilde f}^{\tb{x}}_{L} - \mathbf{\tilde f}^{\tb{x}}_{R}}{\lambda_R - \lambda_L}, \\
    \mathbf{\tilde f}^{\tb{x}, \rm HLL} &=
        \frac{\lambda_R \mathbf{\tilde f}^{\tb{x}}_{L}
        - \lambda_L \mathbf{\tilde f}^{\tb{x}}_{R}
        + \lambda_R \lambda_L \left( \mathbf{\tilde q}_R - \mathbf{\tilde q}_L \right) }
        {\lambda_R - \lambda_L}.
\end{align}
The pressure $P_c$ in the intermediate state can be therefore determined by
\begin{align}
    P_c = P_{cL} = P_{cR} = - \lambda_c \left(\tilde f_{E}^{\rm HLL}\right)^{\tb{x}}
        + \left( \tilde f_{S_{\tb{x}}}^{\rm HLL} \right)^{\tb{x}},
\end{align}
and the conserved quantities in the intermediate $cL/cR$ states can be obtained by
\begin{subequations}
\begin{align}
    &D_{cL/cR} = D_{L/R} \frac{\lambda_{L/R} - v^{\tb{x}}_{L/R}}{\lambda_{L/R} - \lambda_c}, \\
\begin{split}
   & S_{\tb{j}, cL/cR} = \frac{1}{\lambda_{L/R} - \lambda_c} \\
         &\times \left[ S_{\tb{j}, L/R} \left( \lambda_{L/R} - v^{\tb{x}}_{L/R} \right)
        + \left( P_c - P_{L/R} \right) \delta^{\tb{x}}_{\tb{j}} \right],
\end{split} \\
    &E_{cL/cR} = 
        \frac{E_{L/R} \left( \lambda_{L/R} - v^{\tb{x}}_{L/R} \right)
        + P_c \lambda_c - P_{L/R}v^{\tb{x}}_{L/R}}{\lambda_{L/R} - \lambda_c}.
\end{align}
\end{subequations}
Once the numerical flux in the tetrad frame is evaluated,
we can eventually transform it back to the Eulerian observer frame given by
\begin{align}
    \mathbf{f}^x &= - \frac{\beta^x}{\alpha}
\begin{pmatrix}
    D \\
    e^{\tb{i}}{}_j S_{\tb{i}} \\
    E
\end{pmatrix}
+ \sqrt{\gamma^{xx}}
\begin{pmatrix}
   \left( \tilde f_{ D} \right)^{\tb{x}}\\
   e^{\tb{i}}{}_j \left( \tilde f_{ S_{\tb{i}}} \right)^{\tb{x}} \\
   \left( \tilde f_{ E} \right)^{\tb{x}}
\end{pmatrix} .
\end{align}

\subsubsection{Equation of state}
We implement a hybrid equation of state (EOS) in the current version of \texttt{SACRA-2D} where the pressure $P$ and the specific internal energy $\epsilon$ are split into the cold part $P_{\rm cold}/\epsilon_{\rm cold}$ and thermal part $P_{\rm th}/\epsilon_{\rm th}$ as
\begin{align}
    P &= P_{\rm cold} + P_{\rm th}, &
    \epsilon &= \epsilon_{\rm cold} + \epsilon_{\rm th}.
\end{align}
The cold part is described by a phenomenological piecewise polytropic (PWP) EOS \cite{read09} where the realistic EOS is approximated by $n$ pieces of polytrope depending on the transitional density $\rho_i$.
The pressure $P_{\rm cold}$ and the specific internal energy $\epsilon_{\rm cold}$ are parameterized by the rest-mass density $\rho$ as
\begin{align}
&\begin{array}{l}
    P_{\rm cold} = K_i \rho^{\Gamma_i} \\
    \epsilon_{\rm cold} = \displaystyle\frac{K_i}{\Gamma_i-1} \rho^{\Gamma_i -1} + \Delta \epsilon_i
\end{array}, &
    &\text{for $\rho_{i-1} \leq \rho < \rho_{i}$}
\end{align}
where  $i$ runs from $1$ to $n$ with $\rho_0:=0$, 
$K_i$ and $\Gamma_i$ are the polytropic constant and index, respectively,
and $\Delta \epsilon_i$ is determined by imposing the continuity condition on the specific internal energy.

In addition to the cold part, we add the thermal part adopting the gamma-law EOS  given by
\begin{align}
    P_{\rm th} = \rho \left( \Gamma_{\rm th} - 1 \right) \epsilon_{\rm th},
\end{align}
where $\Gamma_\mathrm{th}$ is a constant typically set to $5/3$ in the present work. 

\subsubsection{Recovery of primitive variables}

The recovery of primitive variables $(\rho, u_i, P, \epsilon)$ from conserved variables $\mathbf{q}$ is non-trivial and can only be done numerically.
We implement the primitive recovery procedure for GRHD mentioned in Appendix C of \cite{gale13}.
Here, we briefly outline the implementation of the recovery procedure:
\begin{enumerate}
	\item Evaluate the rescaled quantities that are fixed in the iterations 
	\begin{align}
		r &:= \frac{\sqrt{S_i S^i}}{D}, &		q &:= \frac{E}{D}-1, & k &:= \frac{r}{1+q},
	\end{align}
	\item Set the bounds $[z_-, z_+]$ for the root defined as
	\begin{align}
		z_- &:= \frac{k/2}{\sqrt{1 - k^2 / 4}}, &
		z_+ &:= \frac{k}{\sqrt{1 - k^2}}
	\end{align}
	\item Within the interval $[z_-, z_+]$, we find the root of $f(z) = 0$ with the master function $f(z)$ defined as
	\begin{align} \label{eq:c2p_fn}
		f(z) &:= z - \frac{r}{\hat h(z)},
	\end{align}
	where
	\begin{flalign}
 \begin{split}
		\hat h(z) &:= ( 1 + \hat \epsilon ) ( 1 + \hat a(z) ), \\
		\hat P(z) &:= P( \hat \rho(z), \hat \epsilon(z) ), \\
		\hat a(z) &:= \frac{ \hat P(z)}{ \hat \rho(z) (1 + \hat \epsilon(z)) }, \\
		\hat \rho(z) &:= \frac{D}{\hat w(z)}, \\
		\hat \epsilon(z) &:= \hat w(z) q - z r + \frac{z^2}{1+\hat w(z)}, \\
		\hat w(z) &:= \sqrt{1+z^2}.
\end{split}	
        \end{flalign}
\end{enumerate}

In \texttt{SACRA\_2D}, we numerically solve \cref{eq:c2p_fn} using the Illinois method for bracketing root-finding.
We also set an upper limit for Lorentz factor $w_{\rm max}$ (typically set to be $w_{\rm max} = 100$) and rescale $S_i$ whenever $k$ exceeds certain upper bound following \cite{gale13}.
While this method is robust and always converges to a solution, it does not guarantee that the converged solution satisfies the physical condition.
In particular when the obtained specific internal energy falls below the minimum allowed values of EOS ($\epsilon < \epsilon^{\rm EOS}_{\rm min}$),
we employ an additional primitive recovery using only the conversed density and momentum $(D, S_i)$ together with the zero temperature EOS $h = h_{\rm cold}(\rho)$ following a similar procedure.
\begin{enumerate}
	\item Set the bounds $[z_-, z_+]$ for the root defined as
	\begin{align}
		z_- &:= 0, & z_+ &:= r
	\end{align}
	\item Within the interval $[z_-, z_+]$, we find the root of $\tilde f(z) = 0$ with the master function $\tilde f(z)$ defined as
	\begin{align}
		\tilde f(z) &:= z - \frac{r}{h_{\rm cold}(\hat \rho(z))},
                \quad 
	\end{align}
            \text{where}
    \begin{align}
		\hat \rho(z) &:= \frac{D}{\hat w(z)}  &
		\hat w(z) &:= \sqrt{1+z^2}
	\end{align}
        \item Reset the conversed energy $E$ from primitive variables.
\end{enumerate}

In addition, we impose an artificial atmosphere by defining a lower bound $\rho_{\rm atm}$ and reset the rest-mass density $\rho$ after the primitive recovery whenever it falls below the bound $\rho = \max(\rho, \rho_{\rm atm})$
to maintain stable evolution in the low-density region.
The cutoff density $\rho_{\rm atm}:= \rho_{\max} f_{\rm atm}$ depends on the initial maximum density $\rho_{\max}$ where the auxiliary factor $f_{\rm atm}$ is typically set to be $\leq 10^{-15}$.

\subsection{FMR setting}\label{sec:refinement}

We adopt the fourth-order explicit Runge-Kutta scheme (RK4) in order to evolve the metric function stably \cite{shib16}.
Following the time update scheme in \cite{brug08}, the adaptive time step is employed using the Berger-Oliger algorithm \cite{berg84}.
We allow sub-cycling of time integration starting from level $l_{\rm fix}$ with a time step for each FMR level set to be
\begin{align}
	\Delta t^{(l)} &=
	\begin{cases}
		\Delta t^{(l-1)}, & \text{for } 1 \leq l \leq l_{\rm fix}, \\
		\Delta t^{(l-1)} / 2, &\text{for } l > l_{\rm fix}.
	\end{cases}
\end{align}
The parameter $l_{\rm fix}$ limits the time step in the coarse levels to avoid error induced by over-large $\Delta t$ and reduce the effect from the outer boundary. However, it usually makes no difference practically.
The time step in the finest level $\Delta t^{(L-1)}$ is related to the grid size as
\begin{align}
    \Delta t^{(L-1)} = c_{\rm CFL} \min (\Delta x^{(L-1)}, \Delta z^{(L-1)}),
\end{align}
where the Courant–Friedrichs–Lewy (CFL) factor $c_{\rm CFL}$ is set to be $0.5$ unless stated otherwise.
In \texttt{SACRA-2D}, the buffer zone's $(4+4)$ structure is employed, where
the outer $4$ buffer cells $[N+5, N+8]$ are used for time interpolation between different time slices, 
while the inner $4$ buffer cells $[N+1, N+4]$ act as a buffer zone to dissipate any oscillatory behavior in the time-interpolated values.
This corresponds to the $[1, N+4]$ domain for the first three stages of RK4 time integration and $[1, N]$ for the last stage.

To obtain the buffer zone at the child level from its parent,
we employ the eighth-order Lagrange interpolation for geometric variables
and minmod limiter to reconstruct the primitive hydrodynamics variables $\mathbf{p}$ for the prolongation in space.
For the time interpolation in grid $[N+5, N+8]$, we employ a second-order Lagrange interpolation of three time slices $\{t^{n-1}, t^n, t^{n+1}\}$ of its parent level for time $t^n < t < t^{n+1}$.
Since the buffer zone does not affect the conservation of the hydrodynamics quantities in the FMR setting,
we interpolate the primitive variables $\mathbf{p}$ and construct the conserved variables directly following \cite{ston20} to avoid additional primitive recovery in the buffer zone.
A limiter procedure is also introduced for fluid variables $\mathbf{p}$ following \cite{yama08} to maintain numerical stability, where we modify the time interpolation to first order with time levels $\{t^n, t^{n+1}\}$ if the following relation holds:
\begin{align}
    \left( \mathbf{p}^{n+1} - \mathbf{p}^{n} \right) \left( \mathbf{p}^{n} - \mathbf{p}^{n-1} \right) < 0.
\end{align}

After each time matching step between the child and parent levels,
the grid values are transferred from the child level (fine grid) to the parent level (coarse grid) in the overlap region.
More specifically, the grids $(x,z) \in ([1, N], [1, N])$ in the child level is mapped to $(x,z) \in ([1, N/2], [1, N/2])$ in the parent level.
In this restriction procedure, we employ an eighth-order Lagrange interpolation for the geometric variables,
and the following conservative scheme \cite{ston20}
\begin{align}
    \mathbf{q}^{(l-1)}_{j, k} = \frac{1}{\Delta V^{(l-1)}_{j,k}} 
        \sum_{m = 2j-1}^{2j} \sum_{n = 2k-1}^{2k} \mathbf{q}^{(l)}_{m, n} \Delta V^{(l)}_{m,n},
\end{align}
for the conserved variables $\mathbf{q}$ with $j,k = 1, \cdots, N/2$.
The primitive recovery procedure is carried out afterward to obtain the updated primitive variables $\mathbf{p}$.

As the parent and child levels evolve in a different time step, 
the numerical flux across the refinement boundary becomes inconsistent and introduce 
violation of conservation in mass and angular momentum.
To solve this, we store the numerical flux of all conserved variables at the same refinement boundary for both fine and coarse levels during the time integration.
After each level finishes the sub-cycling and matches time with its parent level,
we correct the conserved variables next to the refinement boundary in the coarse grid by adding the difference of numerical fluxes between the coarse and fine interface \cite{berg89,east12,reis13}.

\subsection{Hybrid Parallelization} \label{sec:parallel}
\texttt{SACRA-2D} is hybrid parallelized by MPI and OpenMP.
We employ a simple domain-based decomposition for MPI parallelization.
Each level is divided into $M_\mathrm{MPI} \times M_\mathrm{MPI}$ blocks of subdomains ($M_{\rm MPI} \times 2M_{\rm MPI}$ in the absence of the mirror symmetry with respect to the $z=0$ plane),
where $M_{\rm MPI}$ is the number of blocks in $x$ and $z$ directions.
The choices of $M_{\rm MPI}$ are limited by the number of grids $N$, which requires $N/M_{\rm MPI}$ to be an even number.
OpenMP further parallelizes the subdomains, with $N_{\rm thr}$ being the number of OpenMP threads in each MPI rank.
The total number of cores required for the simulation is then determined by $M_{\rm MPI}\times M_{\rm MPI} \times N_{\rm thr}$.

\subsection{Diagnostics}

\subsubsection{Constraints, mass, and angular momentum}

We monitor the overall constraint violations by computing the corresponding $L_2$-norm every timestep as
\begin{subequations}
\begin{align}
	|| \mathcal{H} ||_2 &= \int \left( R + K^2 - K_{ij} K^{ij} - 16 \pi E \right) dV, \\
	|| \mathcal{M}_i ||_2 &= \int \left( D_j K^j{}_i - D_i K - 8 \pi J_i \right) dV,
\end{align}
\end{subequations}
where $\mathcal{H}$ and $\mathcal{M}_i$ are the Hamiltonian and momentum constraints, respectively.
Under axisymmetry, the momentum constraints $\mathcal{M}_x$ and $\mathcal{M}_y$ evaluated are effectively
$ \mathcal{M}_\varpi$ and $\mathcal{M}_\phi$ in cylindrical coordinates,  respectively. 

We also compute the total baryon mass and angular momentum as
\begin{subequations}
\begin{align}
	M_b &= \int D \sqrt{\gamma} dV = \int W^{-3} \rho w dV, \\
	J &= \int S_\phi \sqrt{\gamma} dV = \int W^{-3} \rho h w u_y x dV,
\end{align}
\end{subequations}
which should be conserved.
The gravitational mass and angular momentum of the system are also obtained by analyzing the asymptotic behavior of the geometric quantities.

\subsubsection{Extraction of gravitational wave}

We extract gravitational waves from the numerical data using the outgoing component of Newman-Penrose quantity $\Psi_4$ \cite{newm62},
which can be expressed by the electric part $E_{ac}:= C_{abcd} n^b n^d$
and magnetic part $B_{ac}:= \frac{1}{2} C_{abef} \epsilon^{ef}{}_{cd} n^b n^d$ of Weyl tensor $C_{abcd}$ as \cite{yama08,poll11}
\begin{align}
    \Psi_4 = - ( E_{ac} - i B_{ac} ) \bar m^a \bar m^c,
\end{align}
where $\epsilon_{abcd}$ is the covariant Levi-Civita tensor
and $\bar m^a$ is part of the null tetrad $(k^a, l^a, m^a, \bar m^a)$.
Here, $k^a$ and $l^a$ are outgoing and ingoing null vectors, respectively, where $m^a$ is a complex null vector satisfying 
\begin{align}
    - k^a l_a = 1 = m^a \bar m_a.
\end{align}
We construct a set of spherical shells at different radii composed of $N_\theta$ cell-centered grids for $\theta \in [0, \pi]$ 
($\theta \in [0, \pi/2]$ in mirror symmetry) with grid points defined by
\begin{align}
    \theta_j &= \frac{\pi}{N_\theta} \left( j - \frac{1}{2} \right), &
    &\text{for } j = 1,2\cdots, N_\theta,
\end{align}
and extract $\Psi_4$ on the surfaces by Lagrange interpolation.
We further decompose $\Psi_4$ into tensor spherical harmonic modes $(l,m)$ \cite{brug08}
\begin{align}
    \Psi_4^{(l,m)} &= \int \Psi_4 \bar Y^{-2}_{l,m} \left( \theta, \phi \right) d\Omega, 
\end{align}
where $Y^{-2}_{l,m}$ is the spin-weighted spherical harmonic function with $s=-2$.
Due to the axial symmetry, only the $m=0$ modes are extracted with no $\phi$ dependence.
We adopt the accurate Gauss quadrature scheme for the integration following \cite{poll11}.

\subsubsection{Apparent horizon finder}
To identify the presence of a black hole and to diagnose its properties,
we implement an apparent horizon finder in \texttt{SACRA-2D}.
Assuming that the apparent horizon contains the coordinate center $(x,z)=(0,0)$,
the horizon radius $H$ can be represented as a function of polar angle $\theta$ as $r = H(\theta)$.
Under an axisymmetric configuration, the elliptic equation for the radius of the apparent horizon is reduced to one-dimensional.
We essentially employ the same method in \cite{yama08} to solve the equation. 
We note that even if a black hole is located along the $z$-axis different from $z=0$, the finder can find the apparent horizon by simply changing the definition of $\theta$.

Once the radius of the apparent horizon is determined,
we then evaluate its area $A_H$,
and obtain the irreducible mass $M_{\rm irr}$ and the angular momentum $J$ of the black hole as 
\begin{align}
    M_{\rm irr} &= \sqrt{\frac{A_H}{16 \pi}}, &
    J &= \frac{1}{8\pi} \oint_{\mathcal{H}} K_{ab} \phi^a s^b dA,
\end{align}
where $\mathcal{H}$ corresponds to the surface of the apparent horizon,
$\phi^a := (\pa/\pa_\phi)^a$,
and $s^b$ is the unit radial vector normal to $\mathcal{H}$.
As a result, the mass of the black hole can be determined by
\begin{align}
    M = \sqrt{ M_{\rm irr}^2 + \frac{J^2}{4 M_{\rm irr}^2}}.
\end{align}
Once the black hole is formed, we excise the fluid quantities by setting $u_i = 0$ and the rest-mass density to zero for $r \leq H(\theta) / 2$
to avoid any potential numerical instability which may be caused by extreme values of hydrodynamics quantities inside the black hole.

\section{Numerical test}\label{sec:result}
This section presents representative examples of the benchmark test problems with \texttt{SACRA-2D}.
We first examine the metric and GRHD sectors separately with tests considering vacuum spacetime in \cref{sec:vacuum} and fixed background metric in \cref{sec:cowling}, respectively.
The code is then fully tested in \cref{sec:full_test} considering problems that cooperate GRHD in dynamic spacetime.

\subsection{Vacuum spacetime}\label{sec:vacuum}
\subsubsection{Trumpet Black hole}
We first test our metric solver on a stationary spacetime.
Specifically, we consider a non-rotating black hole in the so-called maximal trumpet coordinate,
which is time-independent under BSSN formalism with the puncture gauge.
The analytic solution of the trumpet-puncture black hole is given by \cite{estr73,baum07}
\begin{subequations} \label{eq:trumpet}
\begin{align}
    \alpha &=\sqrt{1 -\frac{2M}{R}+\frac{27 M^4}{16 R^4}}, \\
    \beta^i &=\frac{3\sqrt{3} M^2}{4R^3}x^i,\\
    W &=\frac{r}{R}, \\
    h_{ij} &=0 = K, \\
    \tA_{ij} &=\frac{3\sqrt{3}M^2}{4R^3}\left(\delta_{ij}
    -3 \frac{x^i x^j}{r^2}\right), 
\end{align}
\end{subequations}
where $M$ is the mass of the black hole, $r$ is the radial coordinate,
and $R$ is the areal radius in Schwarzschild metric, which is a function of $r$ written as
\begin{align}
\begin{split}
    r &=\left(\frac{2R + M + \sqrt{4R^2 + 4MR + 3M^2}}{4}\right) \\
    &\times
    \left[\frac{(4+3\sqrt{2})(2R-3M)}
    {8R+6M+3\sqrt{8R^2+8M R+6M^2}}\right]^{1/\sqrt{2}}. 
\end{split}
\end{align}
In this coordinate, $r=0$ corresponds to an areal radius of $R=3M/2$, and the event horizon radius is located at $r\approx 0.78 M$.
To evolve the trumpet data, we use a gauge condition consistent with the staticity of the solution \cite{ruch18} as
\begin{subequations} \label{eq:static_gauge}
\begin{align}
    \pa_t \alpha &= - \alpha \left( 1 - \alpha \right) K, \\
    \pa_t \beta^i &= \frac{3}{4} B^i, \\
    \pa_t B^i &= \pa_t \tG^i - \eta_B B^i,
\end{align}
\end{subequations}
with a damping parameter $\eta_B = 1/M$.
The slicing condition in \cref{eq:static_gauge}, compared to the standard 1+log gauge without advection, gives a lower propagation speed of gauge waves.
We found that the numerical result is closer to the analytical trumpet solution under \cref{eq:static_gauge} due to a smaller effect from the gauge dynamics.
We perform numerical evolution of the trumpet data on a computation domain of $L_{\rm max} = 1600M$ and 11 FMR boxes
with different grid resolutions with $N=64, 128$, and $256$,
which corresponds to the grid spacing of $\Delta x = \Delta z = 0.0244 M, 0.0122 M$ and $0.0061 M$, respectively, on the finest level with $L = 1.56 M$.

\begin{figure}
    \centering
    \includegraphics[width=\linewidth]{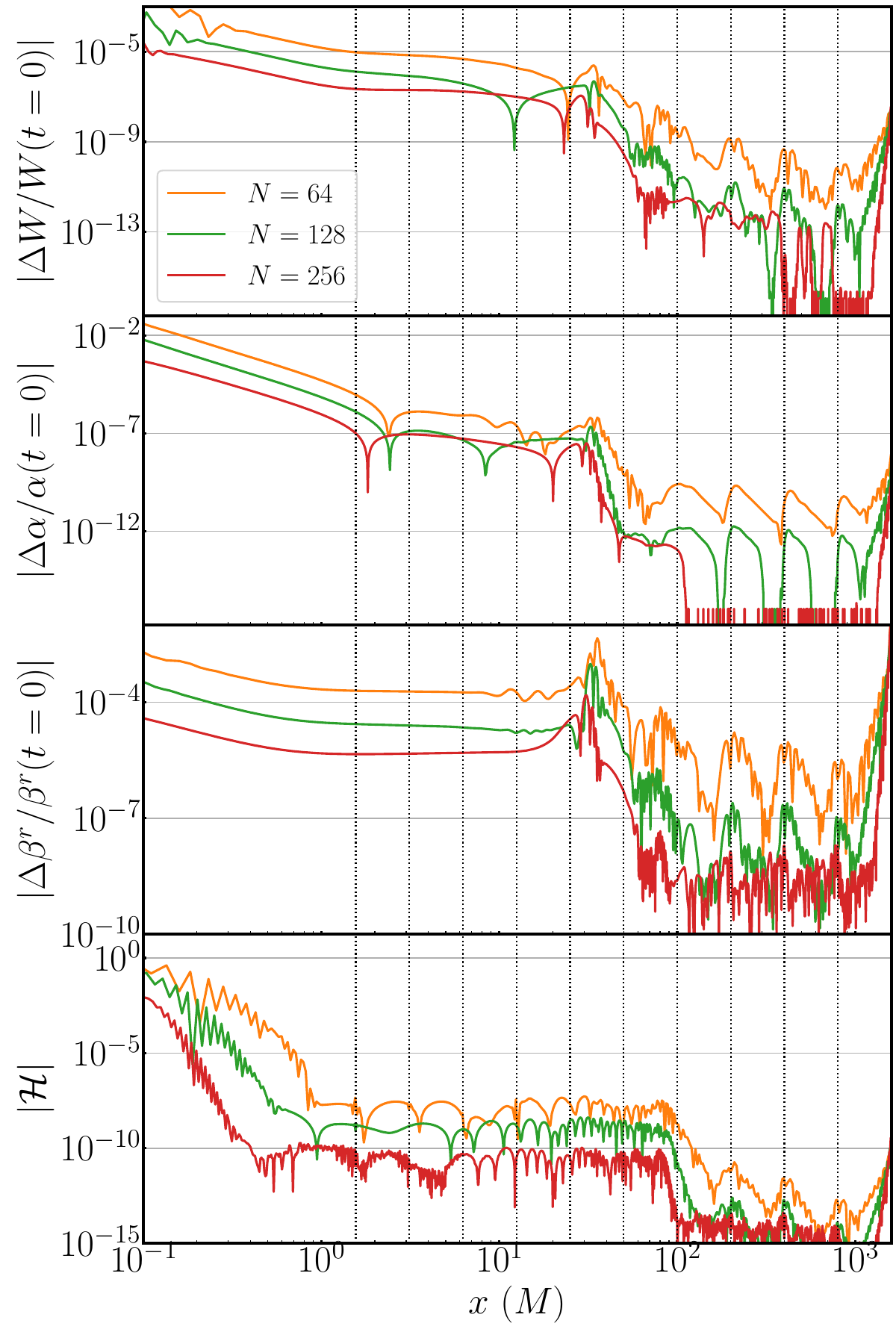}
    \caption{The top three panels show the relativity error of $W:= \psi^{-2}$, lapse function $\alpha$, and shift vector $\beta^r$ along the $x$-axis extracted at $t=195M$ for three different grid resolutions $N=64, 128$, and $256$.
    The bottom panel shows the corresponding absolute violation of Hamiltonian constraints on the same slice.
    The black dotted vertical lines indicate the location of the refinement boundaries.}
    \label{fig:tbh}
\end{figure}

Although the metric variables should remain unchanged analytically under this gauge condition in the trumpet solution,
numerical errors from the finite difference scheme and interpolation across the refinement boundaries will induce deviations from the initial values during the evolution.
We evolve the trumpet data up to $t=195M$ and extract the relative error of $W, \alpha$, and $\beta^r$ as well as the Hamiltonian constraint violation on the $x$-axis at $z=0$ as shown in \cref{fig:tbh}.
Regardless of the resolutions, a spike in relative errors appears at $x\approx 30M$, possibly caused by an outgoing gauge wave.
This could introduce additional noise and induce a loss of convergence \cite{etie14}.
We can recover an expected sixth-order convergence for metric variables in general,
while a roughly fourth-order convergence is found for region $x\lesssim 30M$
where the gauge wave has passed through.
Since we start from time-independent initial data that minimizes gauge dynamics,
the Hamiltonian constraint violations do not contain non-convergent spikes induced by the gauge evolution that appeared in \cite{etie14}, and convergent results are obtained.
In addition, the relative errors and constraint violation show regularly-spaced spikes on a logarithmic space scale in between $x \approx 100M$--$1000M$,
which is a common feature for mesh-refinement structure as the metric variables experience the sudden change in grid spacing across the refinement boundary.

\subsubsection{Spinning Black hole}
To further test our metric solver in a system with non-zero angular momentum,
we evolve a near-extreme-spin black hole with the dimensionless spin parameter $\chi=0.95$.
We adopt the spinning black hole in quasi-isotropic coordinates under a new radial coordinate $r$ introduced in \cite{liu09} defined by
\begin{align}
    r_{\rm BL} &= r \left( 1 + \frac{r_+}{4r} \right)^2,
\end{align}
where $r_{\rm BL}$ is the radial coordinate in Boyer-Lindquist coordinates,
$r_{\pm} = M \pm \sqrt{M^2 - a^2}$ is the Boyer-Lindquist radii of inner $(-)$ and outer $(+)$ horizons of the black hole,
and $M$ and $a$ are the black hole mass and spin, respectively.
The event horizon in this radial coordinate is given by
\begin{align}
    r_h = \frac{1}{4} \left(  M + \sqrt{M^2 - a^2} \right),
\end{align}
which goes to a finite radius $M/4$ when the black hole approaches the maximum spin $a=M$.
This gives better initial data for near-extreme-spin black holes compared to quasi-isotropic coordinates in \cite{kriv98}, in which the coordinate radius of the event horizon drops to zero for $a \rightarrow M$.

The corresponding metric components are written as
\begin{subequations}
\begin{align}
    {}^{(3)} ds^2&= \frac{\Sigma \left( r + \displaystyle\frac{r_+}{4} \right)^2 }{r^3 \left( r_{\rm BL} - r_- \right)} dr^2 + \Sigma d \theta^2
        + \frac{\Xi}{\Sigma} \sin^2 \theta d \phi^2,
\end{align}
\begin{align}
\begin{split}
    K_{r\phi} &= K_{\phi r} = \frac{M a \sin^2 \theta}{\Sigma \sqrt{\Xi \Sigma}} \left( 1 + \frac{r_+}{4 r} \right)
        \frac{1}{\sqrt{r \left( r_{\rm BL} - r_- \right)}} \\
    & \left [ 3 r_{\rm BL}^4 + 2 a^2 r_{\rm BL}^2 - a^4 - a^2 \left( r_{\rm BL}^2 - a^2 \right) \sin^2 \theta \right],
\end{split} \\
\begin{split}
    K_{\theta \phi} &= K_{\phi \theta} = - \frac{2a^3 M r_{\rm BL} \cos\theta \sin^3 \theta}{\Sigma\sqrt{\Xi \Sigma}} 
        \left( r - \frac{r_+}{4} \right) \\
        & \times \sqrt{ \frac{r_{\rm BL} - r_-}{r}},
\end{split}
\end{align}
\end{subequations}
where $\Sigma = r_{\rm BL}^2 + a^2 \cos^2\theta$, $\Delta = r_{\rm BL}^2 - 2M r_{\rm BL} + a^2$,
$\Xi = \left( r_{\rm BL} - a^2 \right)^2 - \Delta a^2 \sin^2\theta$,
and ${}^{(3)} ds^2$ is the spatial line element.

We transform the metric variables to the Cartesian coordinates on the $y=0$ plane and simulate with $a=0.95M$.
The computational domain is set to be $x_{\rm max} = z_{\rm max} = 2048M$ with 10 FMR levels and 
three grid resolutions $N=200, 300$, and $400$,
which correspond to $\Delta x / M= 0.02, 0.0133$, and $0.01$, 
respectively, with the box size $L = 4M$ at the finest level.
We evolve the initial data using the moving puncture gauge of \cref{eq:moving_puncture} with the gauge parameter $\eta_B = 1 / M$.
In this configuration, while the black hole spacetime remains stationary,
the spatial hypersurface will still evolve under the dynamical gauge conditions and eventually approaches the trumpet puncture \cite{hann07,brow08,hann08,brug09,diet14}.

\begin{figure}
    \centering
    \includegraphics[width=\linewidth]{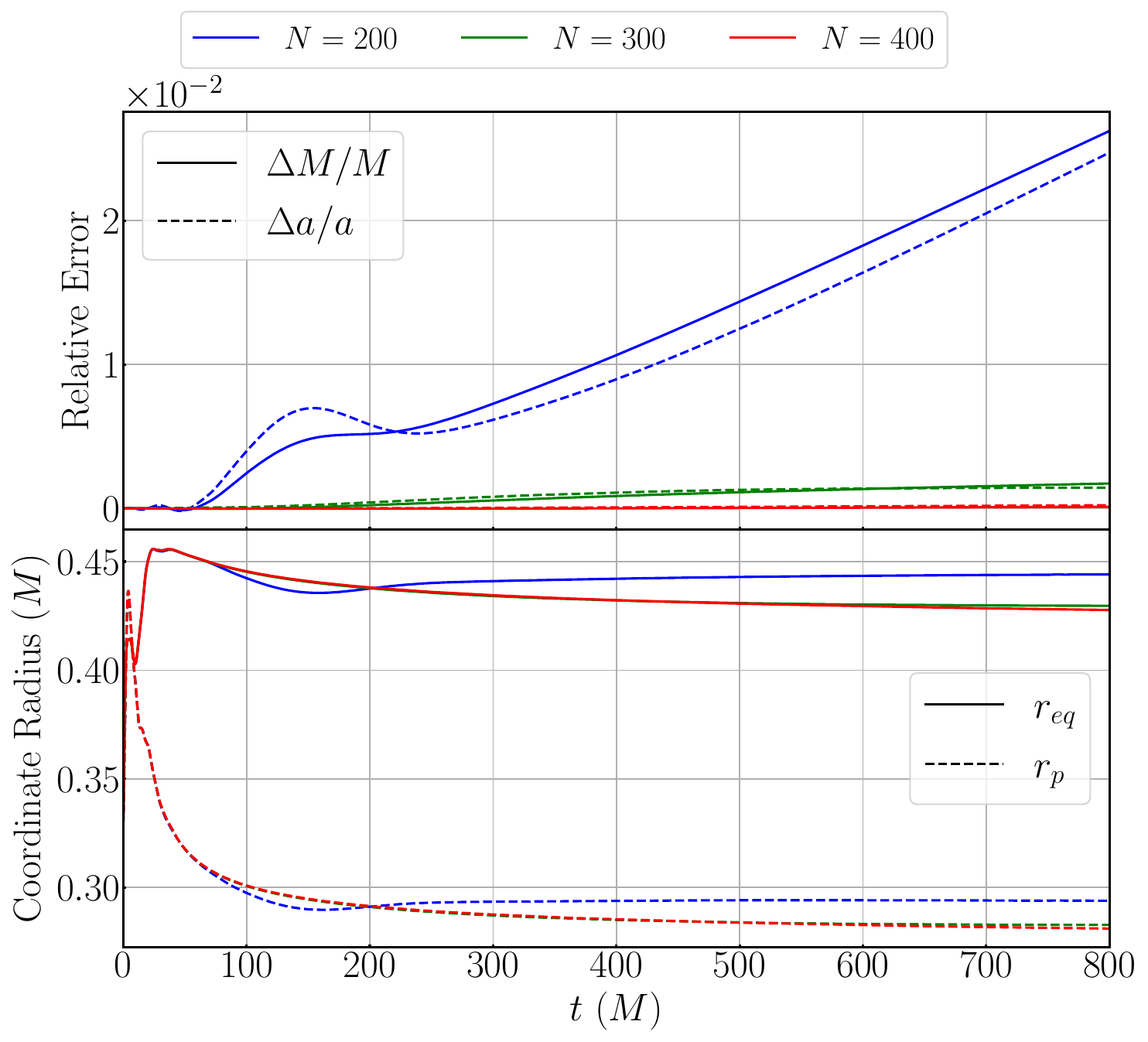}
    \caption{The upper panel shows the relative error of black hole mass (solid) and dimensionless spin (dashed) as functions of time with the initial value of
    $\chi=0.95$.
    The bottom panel shows the evolution of equatorial (solid) and polar (dashed) radii of the apparent horizon in the coordinate radius.
    The initial radius of the apparent horizon is located at $r = 0.328M$.
    The blue, green, red, and cyan lines indicate the result from $N=200, 300$, and $400$, respectively.}
    \label{fig:sbh_a0.95}
\end{figure}

The upper panel of \cref{fig:sbh_a0.95} shows the relative error of the mass $M$ and the spin $a$ of the black hole measured for the apparent horizon. 
As we increase the resolution, the relative error drops and reaches $\sim 10^{-4}$ for the highest resolution with convergence approximately at sixth-order.
On the other hand, the coordinate equatorial $r_{\rm eq}$ and polar $r_{\rm p}$ radii of the apparent horizon evolve under the moving puncture gauge
and eventually approach constant values of 
$r_{\rm eq} = 0.428M$ and $r_{\rm p} = 0.281M$
as shown in the bottom panel of \cref{fig:sbh_a0.95}.
Although both $r_{\rm eq}$ and $r_{\rm p}$ are gauge-dependent quantities,
the values of $r_{\rm eq}$ and $r_{\rm p}$ drop as a result of the hypersurface approaches the trumpet slice of the near-extreme-spin black hole.

\subsubsection{Black hole head-on collision}
To explore the convergence of gravitational waves numerically extracted,
we perform a test simulation of the head-on collision of two non-spinning black holes.
Under axial symmetry, we can set up the Brill-Lindquist initial data \cite{bril63}
which consists of two equal-mass black holes in isotropic coordinates located on the rotational axis ($z$-axis) separated by a distance of $2b$ in the form of
\begin{align}
    \psi &= 1 + \frac{M_{0}}{2 r_{+}} + \frac{M_{0}}{2 r_{-}},
\end{align}
where $M_0$ is constant and
\begin{align}
    r_{\pm} &= \sqrt{x^2 + \left( z \pm b\right)^2 }.
\end{align}
is the radial coordinate distances from the black hole punctures (with $y=0$).

We pick $b=M/2=M_{0}$ following \cite{ruch18} and 
start the simulation under mirror symmetry. Here, $M$ is the total ADM mass of the system, which also defines the unit of length.
In this setup, two black holes are not initially enclosed by the common horizon~\cite{Shibata:1997nc} but merge during the time evolution.
The computational domain is set as $x_{\rm max} = z_{\rm max} = 1024 M$ with 11 FMR levels,
which corresponds to the size $L=1 M$ in the finest box.
We perform the simulations with three grid different resolutions $N=64$ (low), $128$ (med), and $192$ (high) with corresponding resolutions of $\Delta x / M = 0.015625, 0.0078125$, and $0.005208$, respectively, in the finest level.
The non-spinning black holes are released from rest, accelerating toward each other along the polar axis,
and then collide head-on at the origin,
forming a perturbed black hole that promptly rings down to a stationary state by emitting gravitational waves.

\begin{figure}
    \centering
    \includegraphics[width=\linewidth]{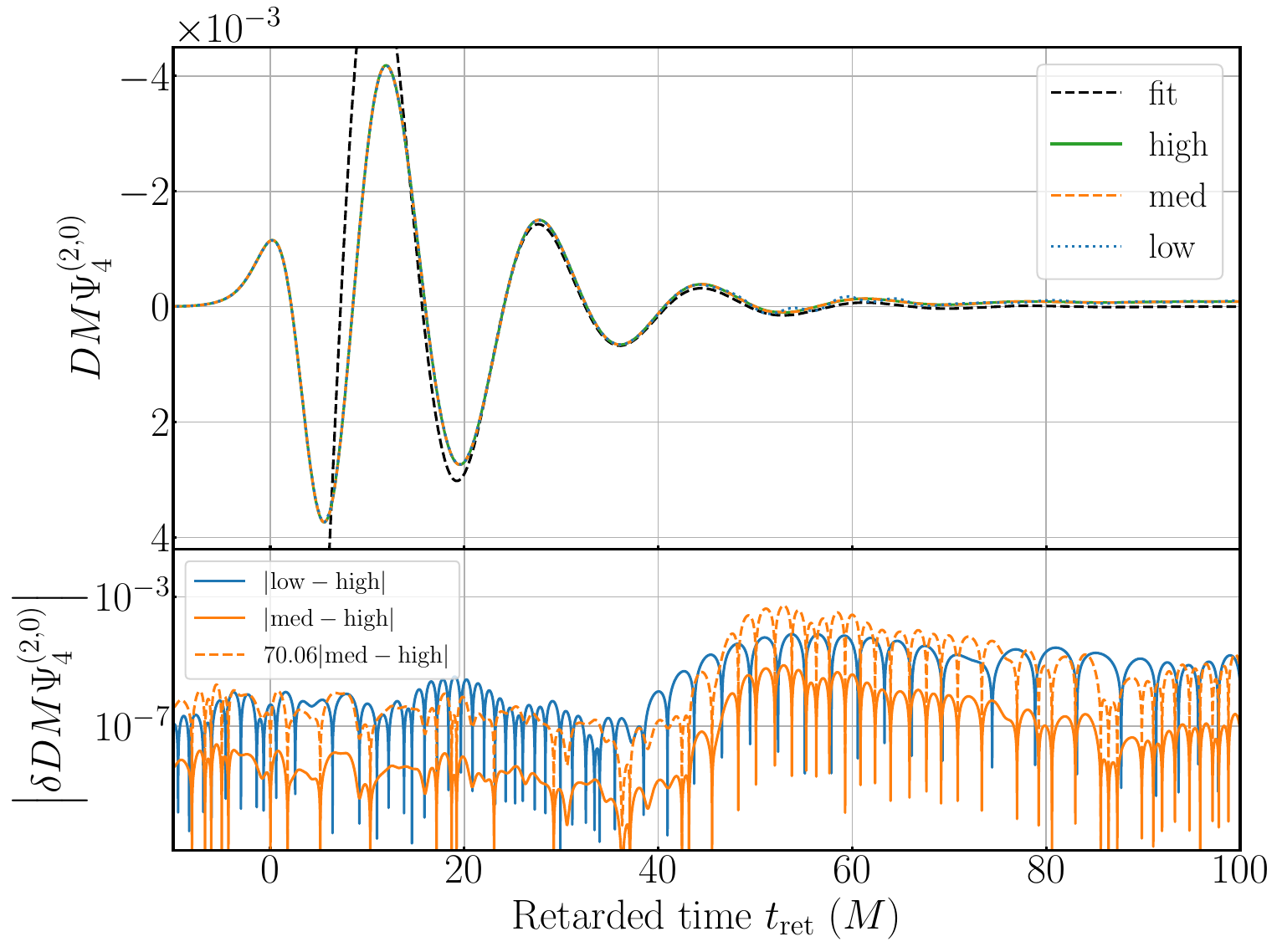}
    \caption{$(l,m)=(2,0)$ mode of $D M \Psi_4$ gravitational waves (top) emitted by the head-on collision of two black holes with extraction radius $r_{\rm ex} = 30M$ in three different grid resolutions.
    The grid spacing $\Delta x / M$ in the finest level from low to high are $0.015625$, $0.0078125$, and $0.005208$, respectively.
    The black dashed line shows the fitted waveform of the analytical ringdown frequency $M\omega \approx 0.3737 - 0.0890i$.
    The bottom panel shows the absolution error of the resultant waveform between different resolutions.}
    \label{fig:headon_psi4}
\end{figure}

\cref{fig:headon_psi4} shows the accompanying gravitational waves signals
extracted at $r_{\rm ex} = 30M$ as a function of retarded time $t_{\rm ret}$ defined by \cite{kiuc17,kiuc20}
\begin{align}
    t_{\rm ret} = t - \left[ D + 2 M \ln \left( \frac{D}{2M} - 1 \right) \right],
\end{align}
where $D \approx r \left[ 1 + M/\left(2r \right)\right]^2$ is areal radius of the extraction sphere.
The resultant ring-down waveform emitted after the merger forms an exponentially damped oscillation with frequency $M\omega \approx 0.3737 - 0.0890i$ \cite{ruch18,bert09} in the dominant $(l,m)=(2,0)$ mode.
The top panel shows the $(l,m)=(2,0)$ mode of $\Psi_4$ in three different grid resolutions,
which are all consistent with the analytical frequency.
In addition, the bottom panel of \cref{fig:headon_psi4} indicates the absolute errors between 
the low ($\Delta x = 0.015625M$) and high ($\Delta x = 0.005208M$) resolutions
as well as the medium ($\Delta x = 0.0078125M$) and high ($\Delta x = 0.005208M$) resolutions as blue and orange solid lines, respectively.
To examine the order of convergence, we scale up the absolute difference between
the medium and high resolutions
by a factor of $(0.015625^6 - 0.005208^6)/ (0.0078125^6 - 0.005208^6)=70.06$
as shown in the orange dashed line,
which agrees approximately with the blue solid line, suggesting the 6th-order convergence in the waveform.
Note that the absolute error of $\Psi_4$ rises, and the convergence is lost for $t_{\rm ret} \gtrsim 40M$.
This is likely caused by the reflection of the outgoing gravitational wave at the refinement boundaries in the coarse domains for which the wavelength of gravitational waves are not well resolved.

\subsection{GRHD with fixed spacetime}\label{sec:cowling}
In this section, we consider test problems with a fixed background metric in both flat Minkowski spacetime and curved spacetime (so-called Cowling approximation),
focusing on the hydrodynamics sector to validate our Riemann solver and reconstruction scheme,
as well as examining the convergence of the hydrodynamics solver.

\subsubsection{one-dimensional shock-tube test}
We carry out a one-dimensional shock-tube test problem following \cite{mart99},
which is commonly used to test the performance of the Riemann solver and reconstruction scheme.
For this test, the cylindrical coordinates in \texttt{SACRA-2D} are converted to the Cartesian coordinates.
Under this setup, the background metric is reduced to the Minkowski flat spacetime with coordinate vector acting as the tetrad basis,
thus allowing us to validate our HLLC solver.
We consider ideal gas law $P=\rho \left(\Gamma-1 \right) \epsilon$ with $\Gamma=5/3$ giving the initial left and right states by
\begin{align}
    \left( \rho, P, v \right) &=
    \begin{cases}
        \left( 10, 40/3, 0 \right) &\text{ for } x<0.5, \\
        \left( 1, 0, 0 \right) &\text{ for } x>0.5,
    \end{cases}
\end{align}
where $v$ is the velocity in the $x$ direction, i.e., $v=u^x/u^t$.
The computational domain is set to be $x \in [0,1]$ with the grid resolution of $N=800$ ($\Delta x = 0.00125$) and no grid refinement.
A third-order PPM scheme is used for reconstruction.

\begin{figure}
    \centering
    \includegraphics[width=\linewidth]{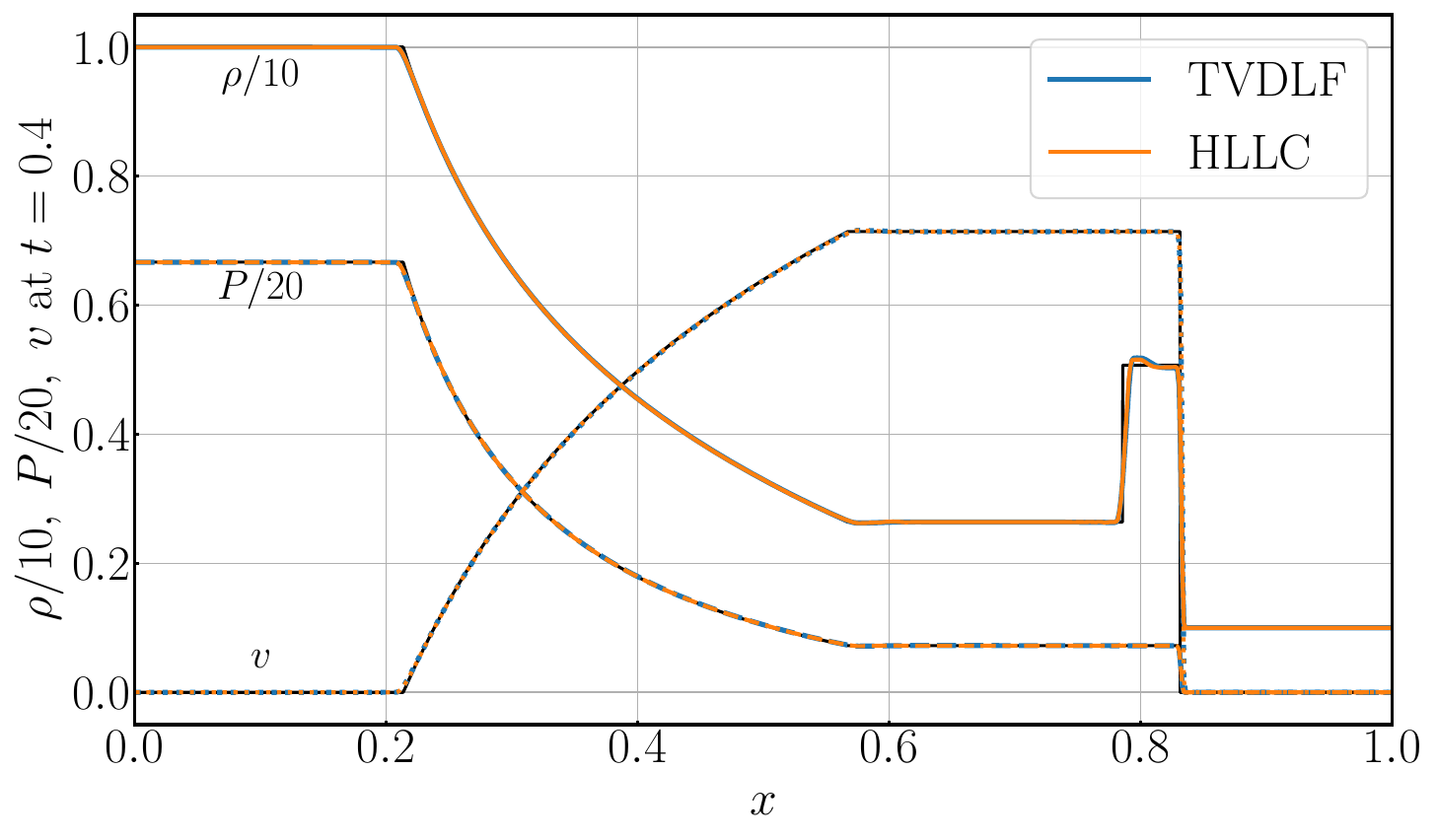}
    \caption{The density (solid), pressure (dashed), and velocity (dotted) profile of one-dimensional shock-tube problem at $t=0.4$ obtained by TVDLF (blue) and HLLC (orange) Riemann solvers.
    The black solid curve indicates the analytical solution.}
    \label{fig:shocktube1d}
\end{figure}

\cref{fig:shocktube1d} shows the profile of the rest-mass density $\rho$, pressure $P$, and velocity $v$ at $t=0.4$ compared to the analytical solutions generated by 
\texttt{RIEMANN} 
\footnote{The open-source program \texttt{RIEMANN} is available \href{https://www.emis.de/journals/LRG/Articles/lrr-2003-7/fulltext.html}{here}.}
\cite{mart99}.
The initial discontinuity at $x = 0.5$ creates left and right propagating shock waves
and forms a contact discontinuity in between,
which is located at $x= 0.786$ for $t=0.4$.
Both TVDLF and HLLC solvers (shown as the blue and orange dots in \cref{fig:shocktube1d}, respectively) can satisfactorily resolve the shocks and contact discontinuity with similar performance,
which is consistent with the result in \cite{kiuc22} when a 3rd-order reconstruction scheme is employed.

\subsubsection{Bondi accretion} \label{sec:bondi}
In this test, we simulate the Bondi accretion \cite{hawl84} consisting of a smooth stationary fluid flow into the black hole that allows us to examine the convergence of hydrodynamics 
and the tetrad formulation for the HLLC solver under a non-trivial spacetime without shocks.
To fit it in the puncture formalism of our code, 
we consider the Bondi solution in the maximal trumpet coordinate of a non-rotating black hole spacetime \cite{mill17},
which does not exhibit coordinate pathology across the event horizon (see \cref{eq:trumpet} for the background metric).
We adopt the same parameters following \cite{whit16} for our setup with an adiabatic index of $\Gamma = 4/3$,
an adiabat of $K=1$, and a critical radius of $r_c = 8M$ with a mass accretion rate $\dot M_{\rm acc} = 0.0848$
where $M$ is the black hole mass.
The hydrodynamics quantities within $r=0.4M$ are fixed as an inner boundary condition.
Six different grid resolutions with $N = 32, 64, 128$, and $256$ are considered for the convergence test.
The computational domain is $x_{\rm max} = z_{\rm max} = 16M$ without mesh refinement.
In addition, we carry out another set of simulations with $3$ refinement levels under the same parameters to test our FMR setting,
which corresponds to a box size $L=4M$ and in the finest box.
Specifically, we examine the convergence of the profile of the rest-mass density $\rho$ by evaluating the $L_{1}$-norm $\epsilon_{\rm L1}$ defined by \cite{whit16}
\begin{align}
    \epsilon_{\rm L1} (\rho) = \frac{\int \left| \rho_{\rm initial} - \rho_{\rm final} \right| \sqrt{-g} dV}{\int \left| \rho_{\rm initial} \right| \sqrt{-g} dV}.
\end{align}

\begin{figure}
    \centering
    \includegraphics[width=\linewidth]{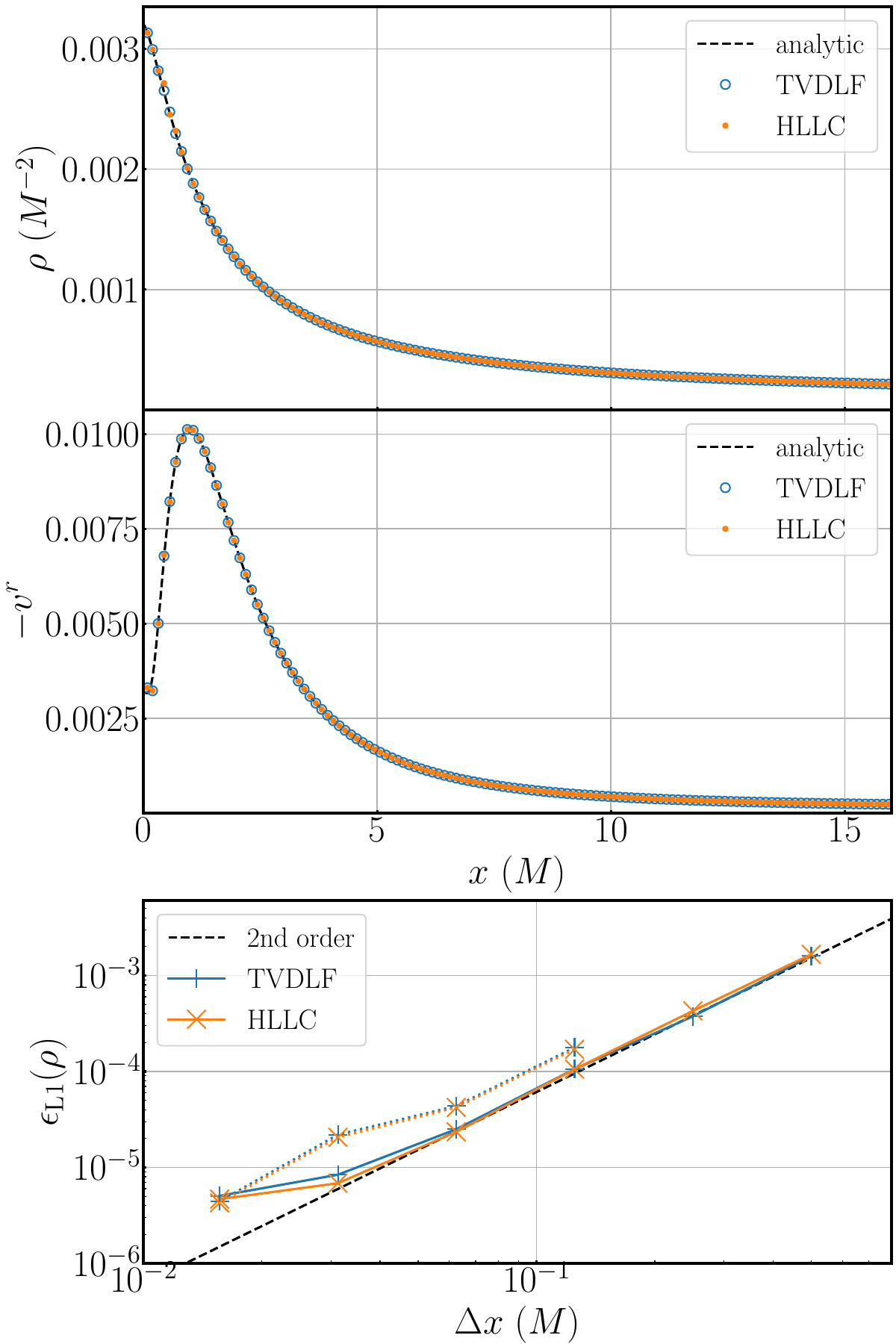}
    \caption{Radial profiles of rest-mass density $\rho$ (top) and radial velocity $-v^r$ (middle) extracted at $t=20M$ with the grid resolution of $N=128$ using TVDLF (blue) and HLLC (orange) Riemann solvers.
    The bottom panel shows the L1 norm of the error in rest-mass density $\epsilon_{\rm L1}(\rho)$ with respective to different grid resolutions in the finest box.
    The solid and dotted lines show the results in the uniform grid setting and the $3$ levels FMR setting, respectively.
    The numerical results are consistent with the second-order convergence (dashed line).}
    \label{fig:bondi}
\end{figure}

The upper and bottom panels \cref{fig:bondi} show the radial profiles of the rest-mass density $\rho$ and the radial velocity $-v^r$ of the Bondi flow, respectively.
The markers show the profiles extracted at $t=20M$ in the resolution of $N=128$ with the TVDLF solver labeled in blue and the HLLC solver labeled in orange,
which agrees approximately with the analytical solution indicated by the black dashed curves.
The bottom panel of \cref{fig:bondi} plots the $L_{1}$-norm of the error of the rest-mass density $\rho$ concerning the different grid spacing.
The TVDLF and HLLC solvers have the same performance due to the smoothness of the accretion flow,
as many other studies have shown (e.g., Refs.~\cite{most14,shib05,kiuc22}).
The result demonstrates an approximate second-order convergence in both solvers regardless of our FMR setting, which is consistent with the accuracy of our implementation of the Riemann solvers.

\subsubsection{Rayleigh-Taylor instability from the modified Bondi flow}
To further validate our HLLC solver and demonstrate its improvement over the TVDLF solver,
we modify the configuration of the Bondi flow to induce Rayleigh-Taylor instability.
Following \cite{xie24}, we change the initial setup within a radius 
$r < 3M [ 1 + 0.05 (\cos ( 80 \theta) + 1) ]$ as
\begin{align}
    \rho &= 0.1\rho_{\rm bondi}, &
    P &= 50 P_{\rm bondi}, &
    u_r &= 0,
\end{align}
where $\rho_{\rm bondi}$ and $P_{\rm bondi}$ are the density and pressure profiles of the Bondi flow in \cref{sec:bondi}, respectively.
This introduces a hot, low-density bubble in the inner region with the perturbed interface.
The hot bubble rises and pushes through the infalling high-density Bondi flow that later on develops the Rayleigh-Taylor-like instability
(or sometimes referred to as the Richtmyer- Meshkov instability).
We employ the same grid setup as in \cref{sec:bondi} with a grid resolution of $N=512$ ($\Delta x = 0.03125M$).

\begin{figure}
    \centering
    \includegraphics[width=\linewidth]{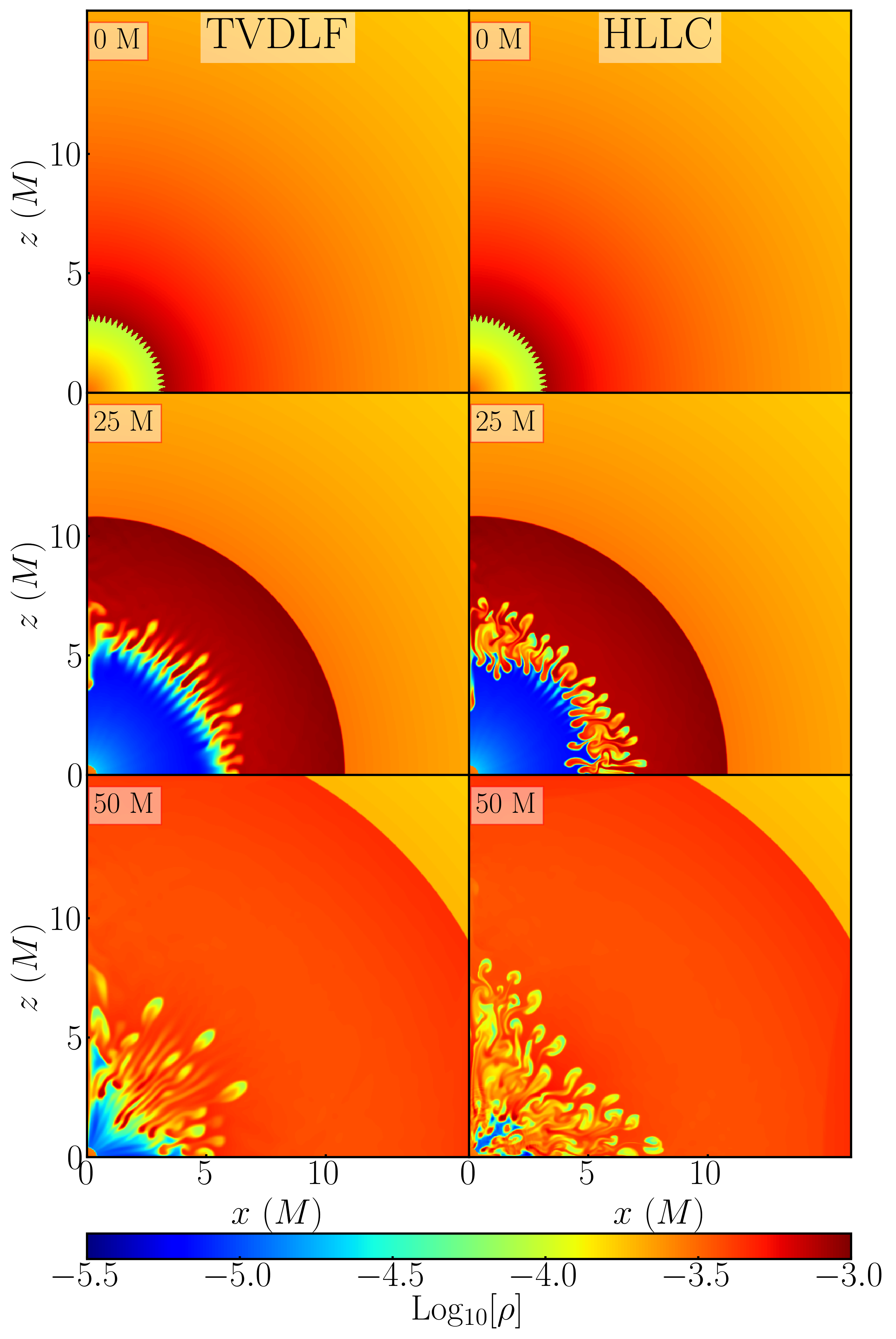}
    \caption{Rest-mass density $\rho$ in the modified Bondi flow with TVDLF (left) and HLLC (right) Riemann solvers.
    The snapshots are extracted at $t/M=0$ (top), $25$ (middle), and $50$ (bottom).}
    \label{fig:modified_bondi}
\end{figure}

\cref{fig:modified_bondi} shows the snapshots of the rest-mass density profile extracted at $t/M=0$ (top row), $25$ (middle row), and $50$ (bottom row).
The left and right columns correspond to the results of the TVDLF solver and the HLLC solver, respectively.
When the hot low-density gas expands and compresses the infalling flow,
instability fingers develop at $t=5M$ and eventually spread inwards at $t=50M$.
The HLLC solver can resolve the Richtmyer-Meshkov instability better than the TVDLF solver,
with the instability finger's fine structure more sharply captured as illustrated in \cref{fig:modified_bondi}.
This demonstrates that the HLLC solver performs better than the TVDLF solver, effectively improving spatial resolution.

\subsection{GRHD with dynamical spacetime}\label{sec:full_test}
In this section, we perform test simulations that solve both hydrodynamics and metric sectors to confirm the full capacity of the code.

\subsubsection{Stable rotating neutron star}\label{sec:stable_ns}

\begin{figure}
    \centering
    \includegraphics[width=\linewidth]{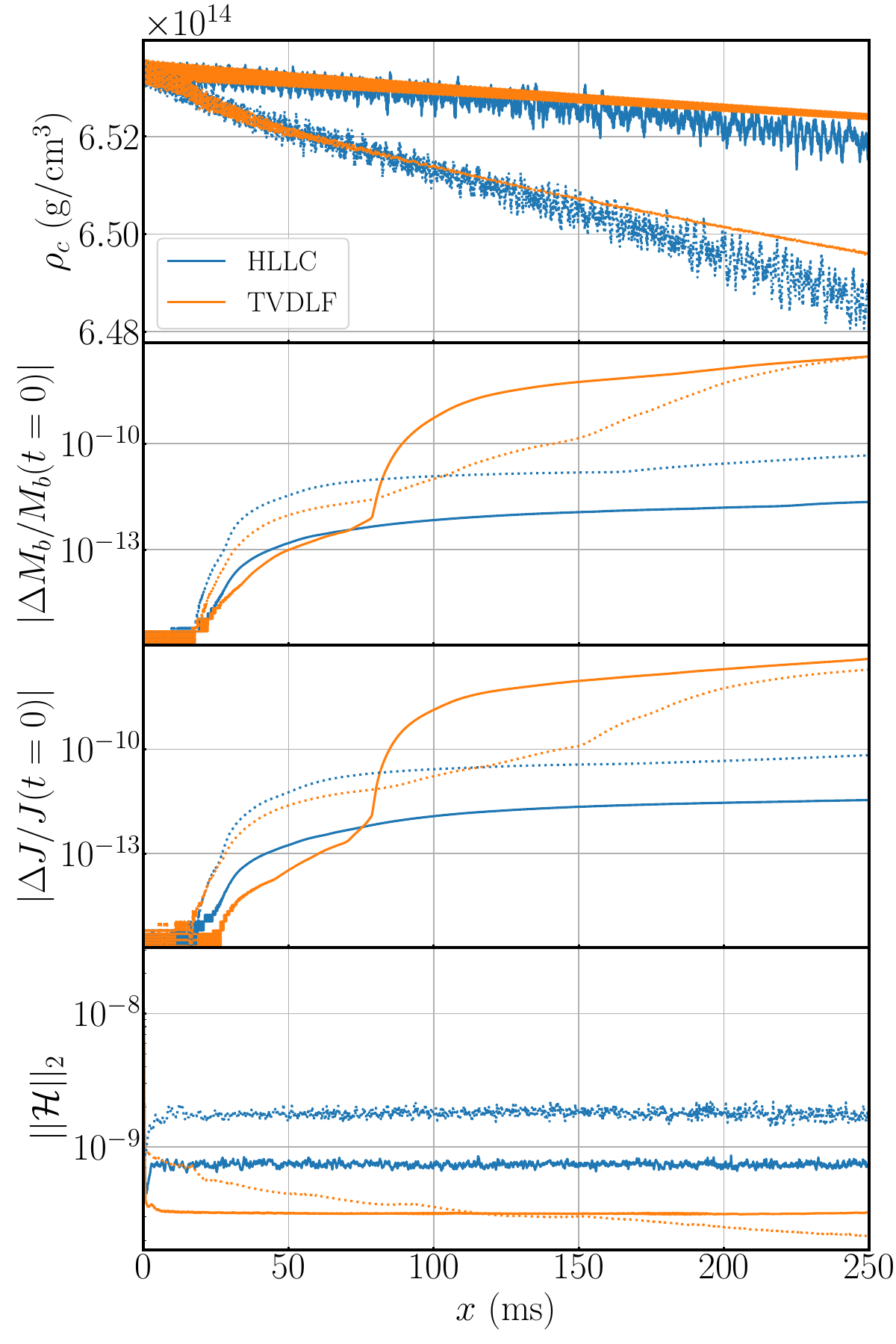}
    \caption{
        From top to bottom, the panels show, respectively, the evolution of central rest-mass density $\rho_c$,
        the relative difference of total baryon mass and angular momentum, 
        and the $L_2$-norm of Hamiltonian constraint violation for the evolution of a rotating neutron star.
        The solid and dotted lines show the results of $N=192$ and $96$, respectively.
    }
    \label{fig:stable_rns_evo}
\end{figure}

\begin{figure}
    \centering
    \includegraphics[width=\linewidth]{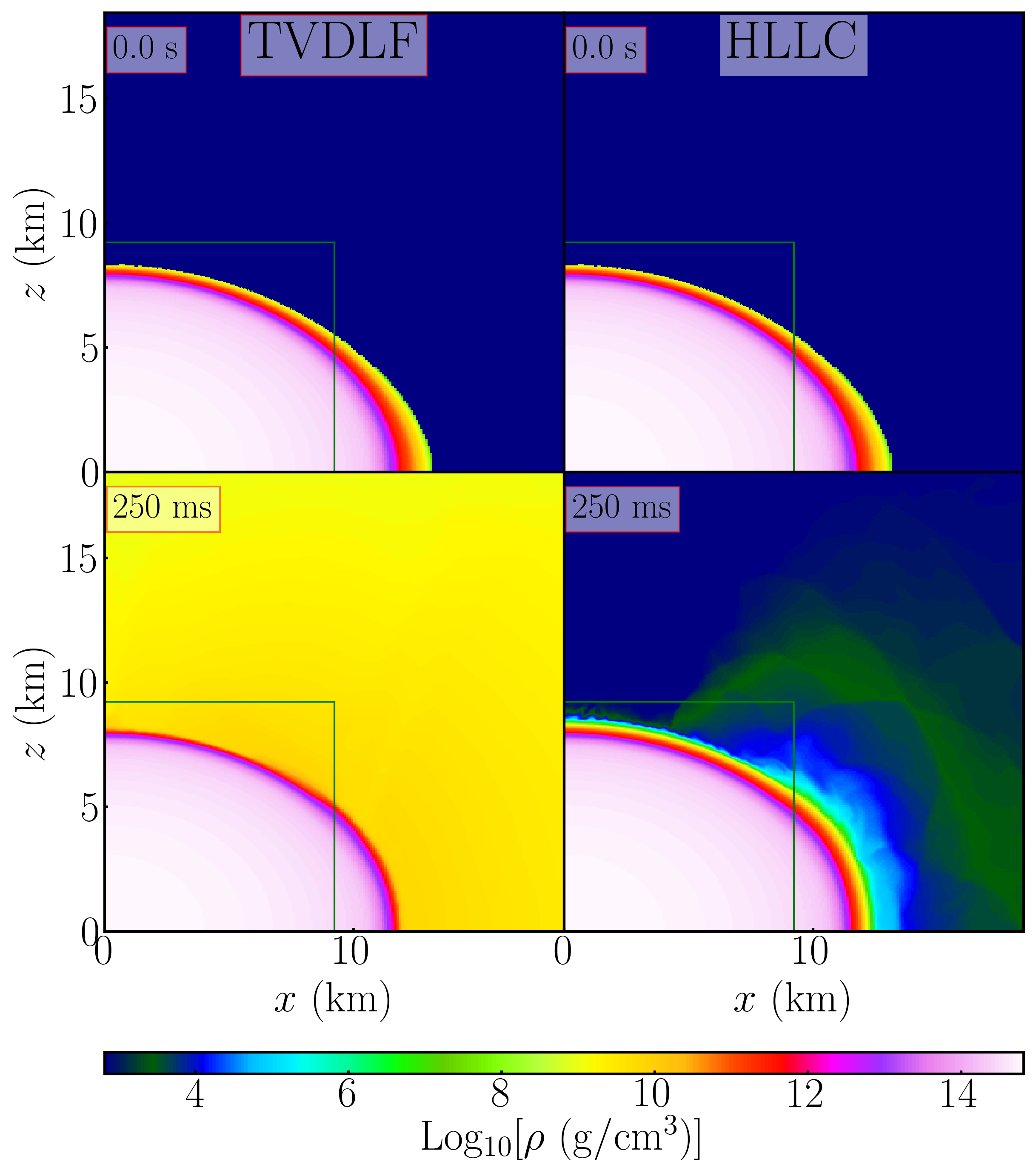}
    \caption{Snapshot of the rest-mass density of the rapidly rotating neutron star with TVDLF (left) and HLLC (right) Riemann solvers for $N=192$.
    The top and bottom rows are extracted at $t=0~{\rm s}$ and $t=250~{\rm ms}$, respectively.
    The green solid lines indicate the boundaries of the FMR levels.}
    \label{fig:stable_rns_snapshot}
\end{figure}

We first evolve a stable rotating neutron star in equilibrium configuration with initial data
constructed by the open-source code \texttt{RNS} \cite{ster95} using the MPA1 EOS \cite{muth87}.
Specifically, a uniformly rotating neutron star with baryon mass $M_b=1.80M_\odot$ and angular momentum $J=1.80M_\odot^2$ is considered
with the ADM mass $M_\mathrm{ADM} = 1.65M_\odot$ and the ratio of rotational kinetic energy to gravitational potential energy $\beta = 0.11$.
At such a high angular frequency $\Omega=6.28~{\rm rad/s}$,
the neutron star is close to its mass shredding limit.
Its shape is flattened to become an oblate spheroid with
the ratio between the coordinate radius at pole $r_{\rm p}$ and equator $r_{\rm eq}$ as $r_{\rm p}/r_{\rm eq}=0.63$.
While such a rapidly rotating neutron star may be subjected to non-axisymmetric $m=2$ bar mode instability \cite{ster98},
it is stable against axial symmetric perturbation.
Hence, maintaining the system stable for a long time in the simulation poses a test problem.

We set the computational domain as $x_{\rm max} = z_{\rm max} = 4726~{\rm km}$ with 10 FMR levels and grid resolution with $N=192$,
which correspond to the size $L=9.23~{\rm km}$ and the resolution $\Delta x = \Delta z = 48~{\rm m}$ in the finest box.
We also carried out the test with a lower resolution $N=96$ as a comparison.
Since the polar and equatorial radii of the neutron star are $r_p = 8.34~{\rm km}$ and $r_{\rm eq} = 13.25~{\rm km}$, respectively,
the refinement boundary of the finest box in this setup cuts through a part of the neutron star as illustrated in the upper panels of \cref{fig:stable_rns_snapshot}. 
This allows us to test the treatment of fluxes and the reconstruction scheme across the refinement boundary with the adaptive time update scheme.
The MPA1 EOS for the cold EOS part and $\Gamma$ thermal law with $\Gamma_{\rm th}=5/3$ are employed.
We perform two sets of simulations using different Riemann solvers and evolve the neutron star up to $t=250~{\rm ms}$,
about $\sim 250$ times the rotational period, which is long enough to examine the quality of the simulation.
In both runs, the mirror symmetry with respect to the equatorial plane is imposed, and the atmosphere factor $f_{\rm atm}$ is set to be $10^{-20}$ and $l_{\rm fix} = 4$.

The top panel of \cref{fig:stable_rns_evo} shows the evolution of the rest-mass density at the center $\rho_c$,
which is maintained throughout the simulation with initial oscillation amplitude $\sim 0.1\%$ and $\lesssim 0.2\%$ shift after $250~{\rm ms}$ for both TVDLF and HLLC Riemann solvers for $N=192$.
The shift converges approximately at the second order.
The oscillation amplitude of $\rho_c$ is noticeably damped out faster for the TVDLF solver than the HLLC solver.
This indicates that the diffusive nature of the TVDLF solver numerically dissipates the oscillation energy,
as another study \cite{kiuc22} also has a similar finding.

The conservation of baryon mass and angular momentum is achieved remarkably well, as shown in the second and third panels of \cref{fig:stable_rns_evo},
with an error of machine precision at $\sim 10~{\rm ms}$,
which validates our treatment of numerical flux across the refinement boundary.
Shortly after that, the relative difference raised to $\alt 10^{-11}$ irrespective of the grid resolution as the matter on the neutron star surface expanded to the atmosphere due to the artificial heating at the surface.
Due to the inability of the TVDLF solver to resolve the contact discontinuity, the effect of the surface heating is much stronger,
creating an artificial outflow and an atmosphere with density $\rho \sim 10^{9}~{\rm g}/{\rm cm}^3$ as shown in the bottom left panel of \cref{fig:stable_rns_snapshot}.
This outflow eventually escapes from the computational domain after $\sim 100~{\rm ms}$ and continuously increases the relative differences of the rest mass and angular momentum to $\sim 10^{-7}$ at the end of the simulation.
On the other hand, these relative differences in the HLLC solver remain $\alt 10^{-10}$ at the end of the simulation irrespective of the grid resolution.
This is because the HLLC solver resolves the surface of the neutron star (i.e., the contact discontinuity) much better than the TVDLF solver,
as shown in the profiles of the rest-mass density $\rho$ in \cref{fig:stable_rns_snapshot}.
At the end of the simulation ($t = 250~{\rm ms}$), the rest-mass density in the atmosphere outside the stellar surface is about $10^4$--$10^5~{\rm g}/{\rm cm}^3$ for the HLLC solver,
which is five orders of magnitude lower than the TVDLF solver, demonstrating a significant improvement in reducing the artificial surface heating.
Note that the structure of the neutron star remains intact across the refinement boundary as shown by the green solid line in \cref{fig:ns_migration_snapshot} without any noticeable numerical artifact despite the adaptive time step treatment. 
This validates our implementation of the FMR scheme.

\subsubsection{Migration of an unstable neutron star}\label{sec:migration}
To further test the nonlinear dynamics of matter and spacetime,
we perform one standard test problem that 
simulates the migration of an unstable neutron star \cite{font02,cord09,bern10,bucc11,cheo21,ng24}.
We construct a Tolman–Oppenheimer–Volkoff (TOV) neutron star in the unstable branch of the mass-radius curve with polytropic EOS $K=100$,
$\Gamma=2$, and central rest-mass density $\rho_c = 8\times 10^{-3}$ (in the unit of $c=G=M_\odot=1$).
Since the unstable branch has a smaller absolute value of the binding energy than its stable companion for this EOS,
the unstable neutron star would migrate to the corresponding stable state with the same baryon mass in the simulation.
The computational domain is set to be $x_{\rm max}=z_{\rm max} = 2215~{\rm km}$ in the grid resolution of $N=128$ with 9 FMR levels,
which corresponds to the box size $L=8.65~{\rm km}$ with the grid spacing $\Delta x = 67.6~{\rm m}$ in the finest level.
We carry out two sets of simulations for this system under the mirror symmetry, one using the $\Gamma$ thermal law with $\Gamma_{\rm th} = 2$,
and another adopting the "adiabatic" EOS \cite{font02}
which neglects the thermal part and enforces zero temperature by discarding the energy equation for $E$.
For both runs, the HLLC Riemann solver is employed with atmosphere factor $f_{\rm atm} = 10^{-15}$ and $l_{\rm fix} = 4$.

\begin{figure}
    \centering
    \includegraphics[width=\linewidth]{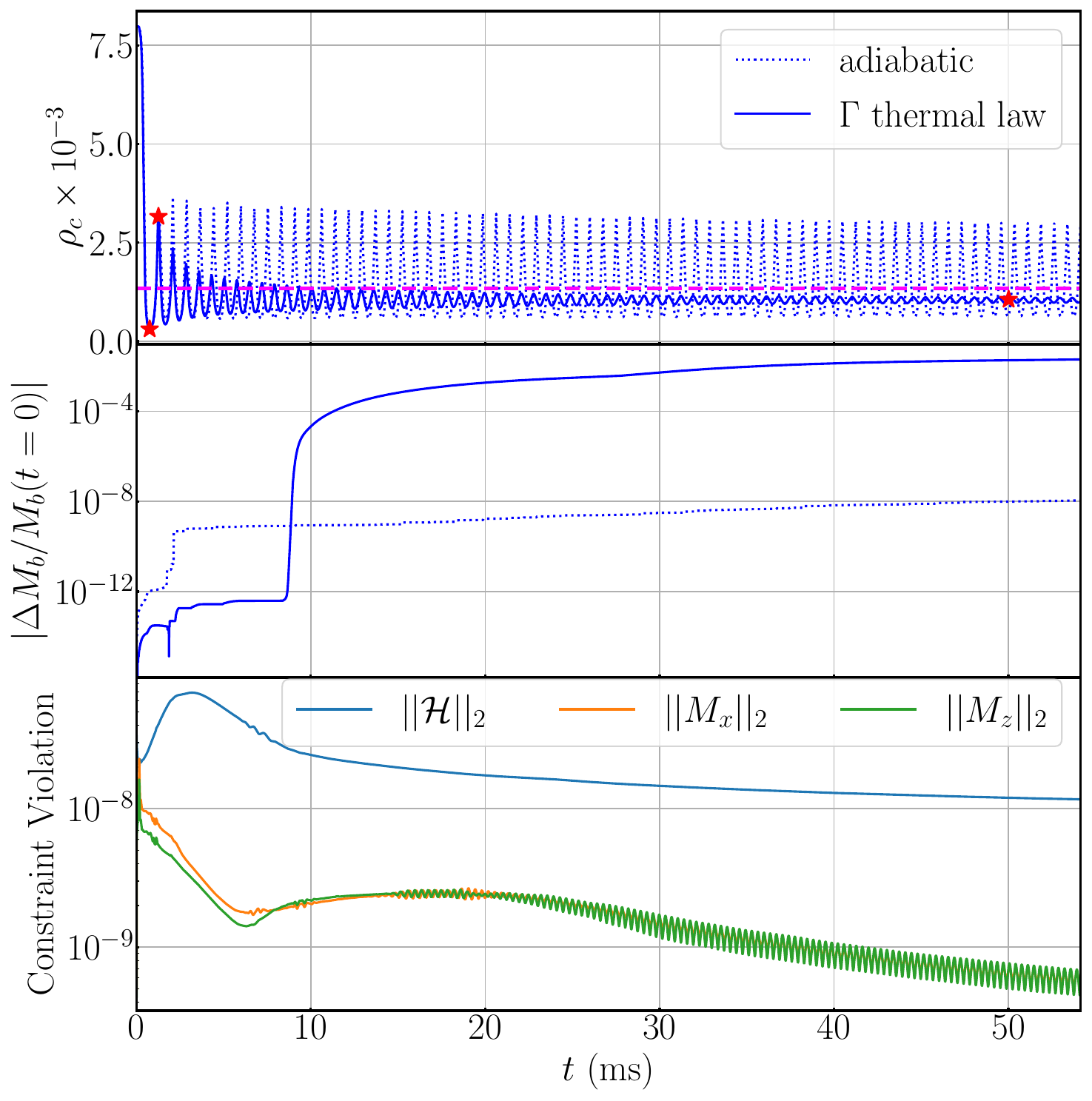}
    \caption{The top and middle panels show, respectively, the evolution of the central rest-mass density $\rho_c$ and the relative difference of the total baryon mass $M_b$ as a function of time with the $\Gamma$ thermal law EOS (solid) and adiabatic EOS (dotted).
    The magenta dashed line on the top panel indicates central rest-mass density $\rho_s = 1.346 \times 10^{-3}$ of the neutron-star model that lies on the stable branch of the mass-radius curve with the same baryon mass.
    The red star markers specify the time extracted for the snapshots shown in \cref{fig:ns_migration_snapshot}.
    The bottom panel shows the $L_2$-norm of constraint violations for the $\Gamma$ thermal law EOS model.}
    \label{fig:ns_migration_evo}
\end{figure}

\begin{figure}
    \centering
    \includegraphics[width=\linewidth]{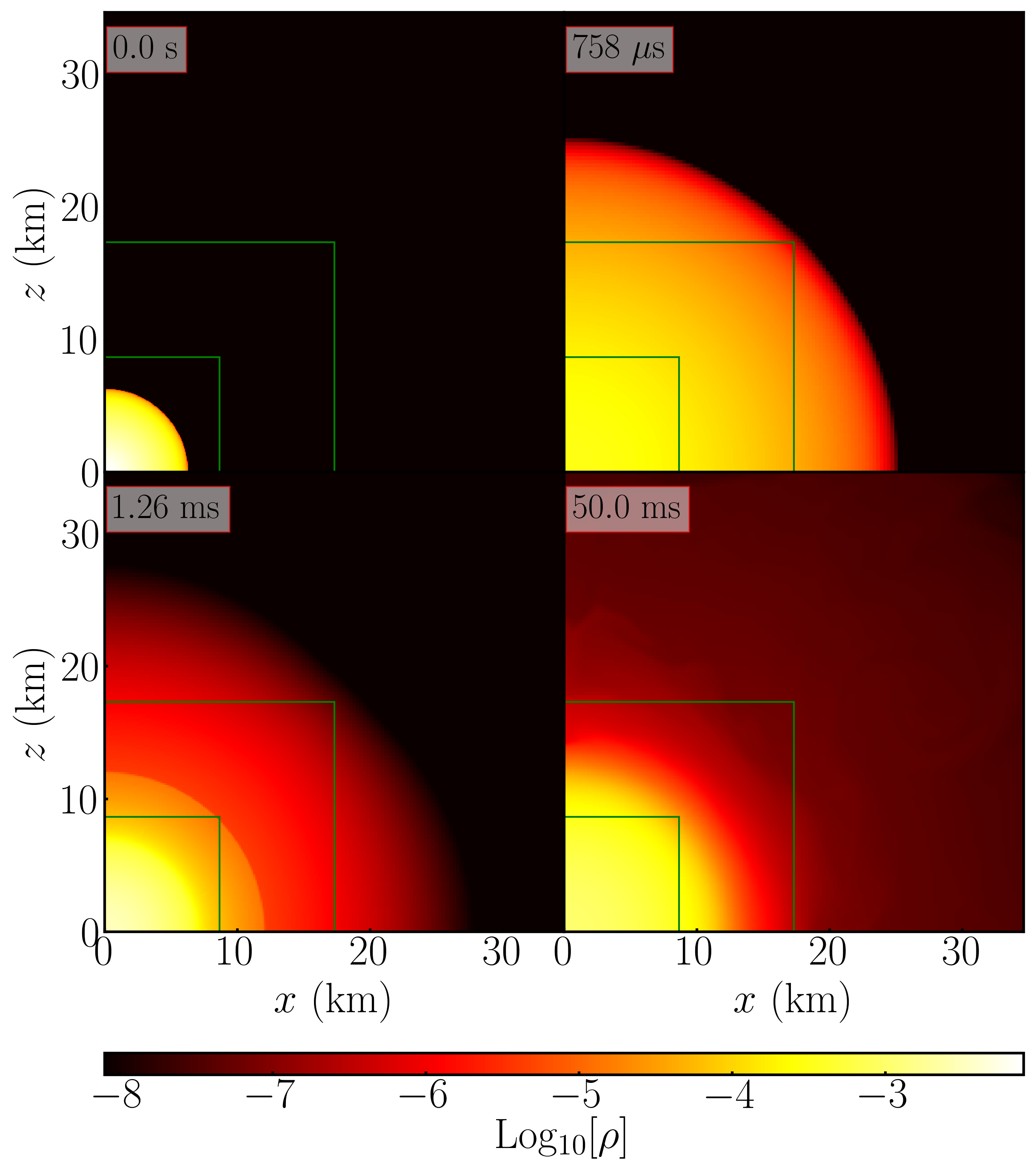}
    \caption{Profiles of the rest-mass density $\rho$ of the unstable neutron star in the migration test extracted at
    $t=0~{\rm s}$ (top left),
    $758~\mu{\rm s}$ (top right),
    $1.26~{\rm ms}$ (bottom left), and
    $50.0~{\rm ms}$ (bottom right) with the $\Gamma$ thermal law EOS employed.
    The green solid lines indicate the boundaries of the FMR levels.}
    \label{fig:ns_migration_snapshot}
\end{figure}

The top panel of \cref{fig:ns_migration_evo} shows the evolution of the central rest-mass density $\rho_c$ as a function of time for the $\Gamma$ thermal law EOS and adiabatic EOS shown in blue solid and dotted lines, respectively.
The red markers indicate the time extracted for the profiles of rest-mass density shown in \cref{fig:ns_migration_snapshot},
and the horizontal magenta dashed line denotes the central rest-mass density $\rho_s = 1.346 \times 10^{-3}$ of the neutron star model on the stable branch with the same baryon mass.
Here, we first focus on the result from the $\Gamma$ thermal law model.
At the start, the neutron star with an initial radius of $6.31~{\rm km}$ immediately swells and tries to migrate to the corresponding stable state.
The central rest-mass density $\rho_c$ rapidly declines and drops below $\rho_s$ to
reach its first minimum at $t=758~\mu{\rm s}$,
with the stellar radius stretching to almost four times larger.
The neutron star compresses and shrinks subsequently until $\rho_c$ reaches its maximum at $t=1.26~{\rm ms}$,
then expands again and hits the infalling matter, forming a shock wave that propagates outwards and ejects a small amount of matter with a high velocity from the stellar surface to the atmosphere. 
We find that the highest velocity of the ejecta is $0.98c$ (Lorentz factor of $\sim 5$), and our code can follow the motion of such a high-velocity component.
The matter ejected by this ejection process eventually leaves the computation domain, which accounts for the sudden rise in the relative difference of total baryon mass, as shown in the middle panel of \cref{fig:ns_migration_evo}.

Nonetheless, the oscillations of $\rho_c$ are gradually damped out in the $\Gamma$ thermal law EOS since the kinetic energy is dissipated to thermal energy through shock heating.
After $t \approx 50~{\rm ms}$, the neutron star approximately settles to a new stable state with $\rho_c$ slightly below $\rho_s$.
In contrast, the neutron star under the adiabatic EOS oscillates with a nearly constant amplitude in the absence of thermal dissipation as the energy converts back and forth between gravitational binding energy and kinetic energy,
which is consistent with the result in \cite{font02}.
This also explains the lower relative difference of total baryon mass since less matter is ejected without shock heating.

We also monitor the $L_2$-norm of the constraint violation of the system as shown in the bottom panel of \cref{fig:ns_migration_evo}.
The Hamiltonian and momentum constraints are well under control, with violations damped out and stabilized under the Z4c prescription. 

\subsubsection{Migration of an unstable rotating neutron star}

\begin{figure}
    \centering
    \includegraphics[width=\linewidth]{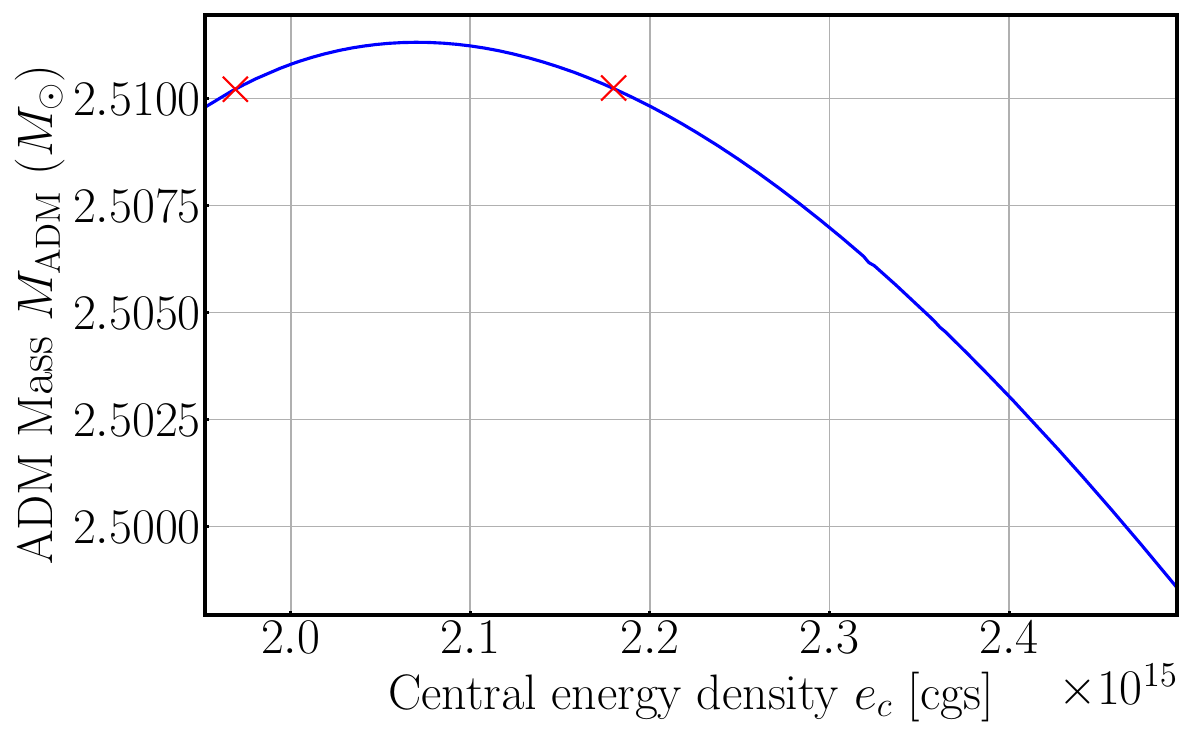}
    \caption{
        The mass versus energy-density ($M_{\rm ADM}$-$\epsilon_c$) curve of the uniformly rotating neutron star of $J=1.800M_\odot^2$ with the MPA1 EOS.
        The red crosses indicate the models selected for the simulations.
        The left marker lies on the stable branch, while the right is on the unstable branch.
        The turning point is located at $e_c = 2.0695 \times 10^{15}~[{\rm cgs}]$ with $M_{\rm ADM}=2.5113M_\odot$.
    }
    \label{fig:MPA1_J1.8_seq}
\end{figure}

\begin{figure}
    \centering
    \includegraphics[width=\linewidth]{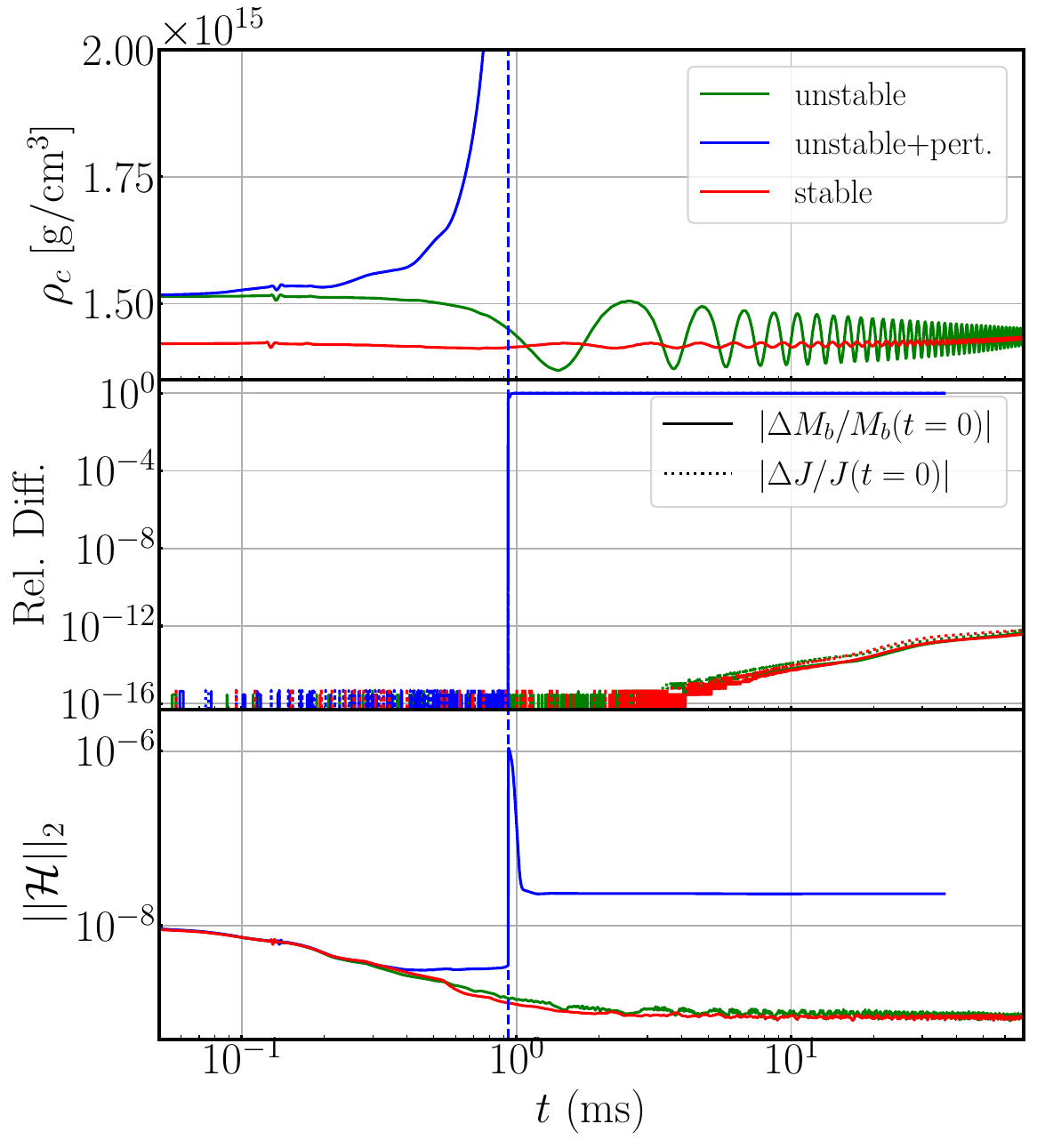}
    \caption{
        The panels show the evolution of central rest-mass density $\rho_c$ (top),
        the relative difference of total baryon mass $M_b$ and angular momentum $J$ (middle),
        and the $L_2$-norm of Hamiltonian constraint violation $||\mathcal{H}||_2$ (bottom) as functions of time.
        The green, blue, and red solid curves illustrate the results from the unstable model, unstable plus initial perturbation model, and stable model, respectively.
        The blue vertical dashed line indicates the black hole formation time $t_{\rm AH} = 0.931~{\rm ms}$ for the unstable plus initial perturbation model.
    }
    \label{fig:rns_migration_evo}
\end{figure}

In this test, we simulate the rotating neutron stars that are very close to the turning point of the mass versus energy density 
($M_{\rm ADM}$-$e_c$) curve to examine the performance of \texttt{SACRA-2D}.
We consider uniformly rotating neutron stars in both the stable and unstable branches indicated as the red crosses in \cref{fig:MPA1_J1.8_seq}
with the same baryon mass $M_b=3.050M_\odot$ and angular momentum $J = 1.800M_\odot^2$ constructed by \texttt{RNS} using the MPA1 EOS.
The parameters of the stable and unstable models are listed in \cref{tab:rns}.
Note that the ADM mass at the turning point $M_{\rm ADM}=2.5113M_\odot$ is only $\approx 0.04\%$ higher than the models we selected,
which poses a challenge for numerical codes in resolving the model accurately.

\begin{table}
\centering
\caption{The parameters of the stable and unstable models used in the simulations.}
\begin{tabular}{
        >{\centering}p{0.50\columnwidth} | 
	>{\centering}p{0.20\columnwidth} 
        >{\centering\arraybackslash}p{0.20\columnwidth} 
    }
 \hline \\[-.8em]
 Models & stable & unstable \\
 \hline \hline \\[-.8em]
 Central energy density             & \multicolumn{1}{c}{\multirow{2}{*}{1.9691}} & \multicolumn{1}{c}{\multirow{2}{*}{2.1798}} \\
 $e_c \times 10^{15}~[{\rm cgs}]$   & & \\
 \hline
  Central rest-mass density             & \multicolumn{1}{c}{\multirow{2}{*}{1.4219}} & \multicolumn{1}{c}{\multirow{2}{*}{1.5151}} \\
 $\rho_c \times 10^{15}~[{\rm cgs}]$   & & \\
 \hline
 ADM mass       & \multicolumn{1}{c}{\multirow{2}{*}{2.510220}} & \multicolumn{1}{c}{\multirow{2}{*}{2.510248}} \\
 $M_{\rm ADM}~[M_\odot]$  & & \\
 \hline
 Baryon mass & \multicolumn{1}{c}{\multirow{2}{*}{3.0500}} & \multicolumn{1}{c}{\multirow{2}{*}{3.0500}} \\
 $M_b~[M_\odot]$ & & \\
 \hline
 Angular frequency & \multicolumn{1}{c}{\multirow{2}{*}{4.6795}} & \multicolumn{1}{c}{\multirow{2}{*}{4.7925}} \\
 $\Omega\times 10^3 ~[{\rm rad}/s]$ & & \\
 \hline
 Equatorial radius & \multicolumn{1}{c}{\multirow{2}{*}{7.425}} & \multicolumn{1}{c}{\multirow{2}{*}{7.215}} \\
 $r_{\rm eq}~[{\rm km}]$ & & \\
 \hline
 Axial ratio &\multicolumn{1}{c}{ \multirow{2}{*}{0.940}} & \multicolumn{1}{c}{\multirow{2}{*}{0.937}} \\
 $r_{\rm p}/r_{\rm eq}$ & & \\
 \hline
\end{tabular}
\label{tab:rns}
\end{table}

Since the unstable model has a higher ADM mass and, hence, a smaller absolute value of binding energy than the stable one,
the unstable neutron star can migrate to the stable configuration similar to the non-rotating case in \cref{sec:migration},
given that the initial numerical perturbation is tiny.
On the other hand, to examine the performance of \texttt{SACRA-2D} for the black hole formation,
we consider an additional run by introducing a small ingoing radial velocity inside the unstable neutron star in forms
\begin{align}
    u_x &= - 5 \times 10^{-3} x / R_{\rm eq}, &
    u_z &= - 5 \times 10^{-3} z / R_{\rm eq}, 
\end{align}
as an initial perturbation to initiate the collapse, where $R_{\rm eq}$ is the star's equatorial coordinate radius.
The computational domain is set to be $x_{\rm max}=z_{\rm max} = 4431~{\rm km}$ in the grid resolution of $N=256$ with 10 FMR levels,
which corresponds to the box size $L=8.65~{\rm km}$ with the grid spacing $\Delta x = 33.8~{\rm m}$ in the finest level.
We carry out three sets of simulations in total under the mirror symmetry,
including one for stable neutron star, one for unstable neutron star without initial perturbation, and one for unstable neutron star with initial perturbation.
We employ the $\Gamma$ thermal law EOS with $\Gamma_{\rm th}=5/3$ and the HLLC Riemann solver for all runs with atmosphere factor $f_{\rm atm} = 10^{-20}$ and $l_{\rm fix} = 4$.
We perform the simulations up to $70~{\rm ms}$, about $\sim 50$ cycles of rotation, for models that do not undergo gravitational collapse.
If the neutron star collapses, we end the run at $30~{\rm ms}$ after the black hole is formed.

\begin{figure}
    \centering
    \includegraphics[width=\linewidth]{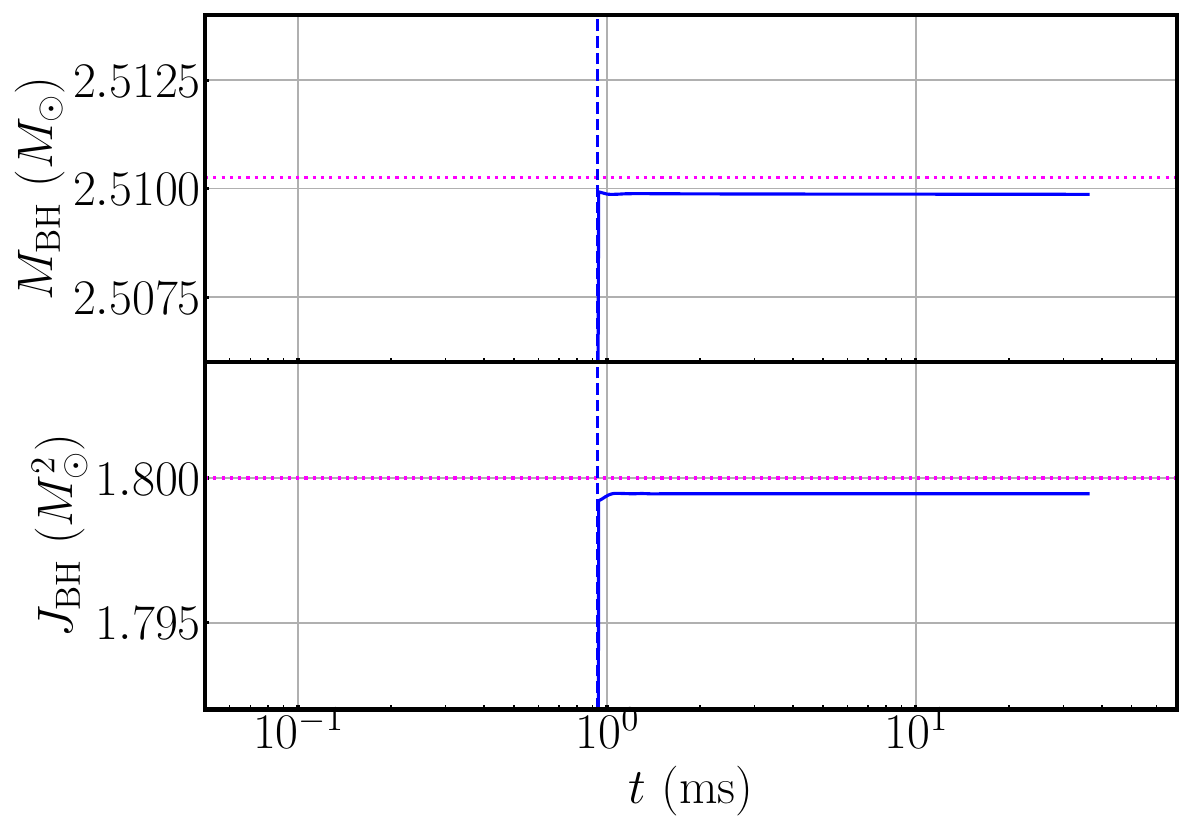}
    \caption{
        The panels show the mass $M_{\rm BH}$ (top) and angular momentum $J_{\rm BH}$ (bottom) of the remnant black hole after collapse for unstable plus initial perturbation model.
        The blue vertical dashed line indicates the black hole formation time $t_{\rm AH} = 0.931~{\rm ms}$.
        The magenta horizontal dotted lines denote the ADM mass $M_{\rm ADM}=2.510248M_\odot$ and the angular momentum $J=1.800M_\odot^2$ of the rotating neutron star obtained from \texttt{RNS}.
    }
    \label{fig:rns_migration_bh}
\end{figure}

\begin{figure}
    \centering
    \includegraphics[width=\linewidth]{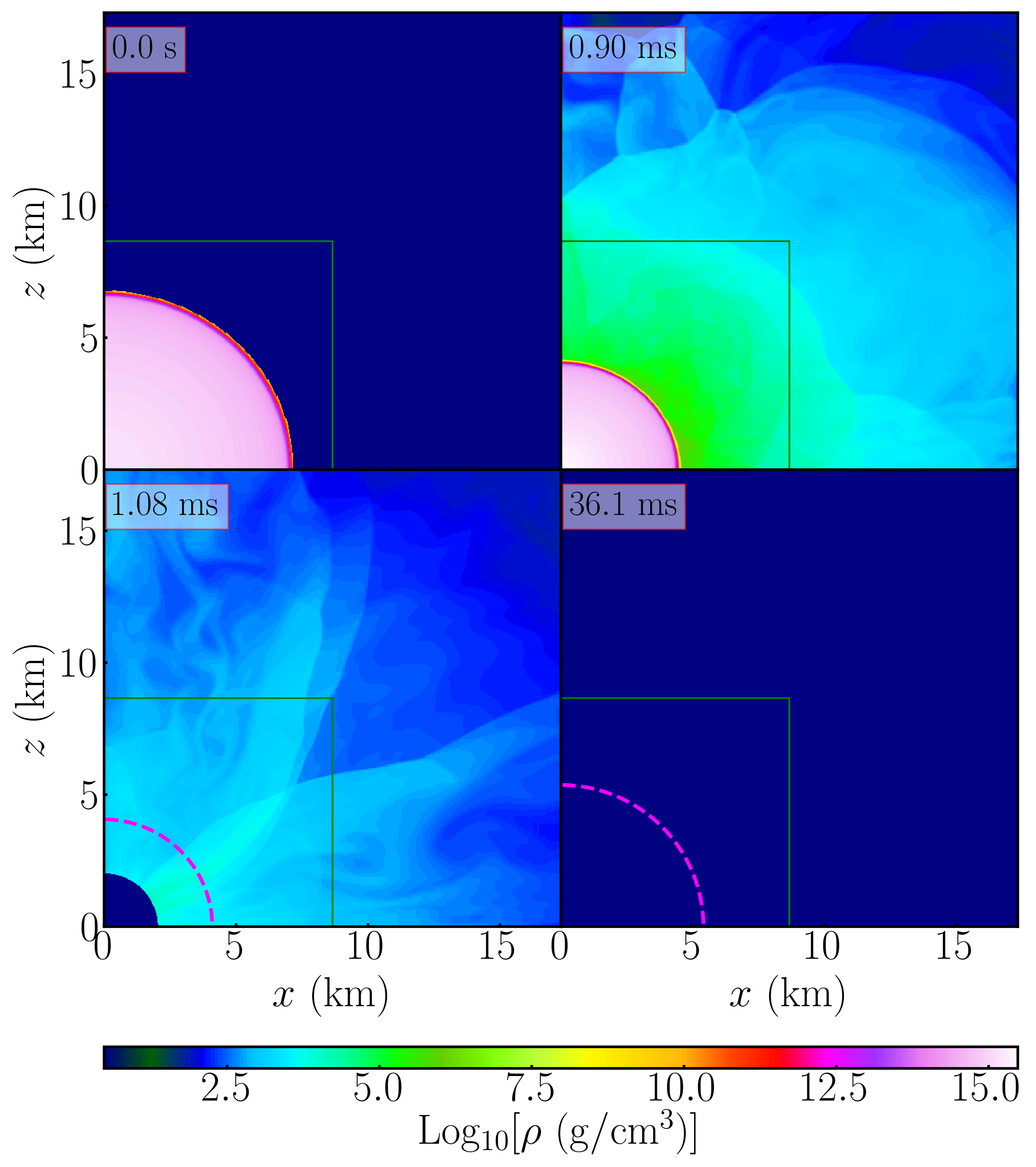}
    \caption{The profiles of the rest-mass density $\rho$ of the collapse of an unstable rotating neutron star extracted at
    $t=0~{\rm s}$ (top left),
    $0.90~{\rm ms}$ (top right),
    $1.08~{\rm ms}$ (bottom left), and
    $36.1~{\rm ms}$ (bottom right).
    The green solid lines indicate the boundaries of the FMR levels,
    and the magenta dashed curves in the bottom panels denote the apparent horizon surface of the black hole.}
    \label{fig:rns_collapse_snapshot}
\end{figure}

\cref{fig:rns_migration_evo}, from top to bottom, shows the evolution of the central rest-mass density $\rho_c$, 
the relative differences of total baryon mass $M_b$ and angular momentum $J$,
and the $L_2$-norm of the Hamiltonian constraint violation $||\mathcal{H}||_2$ as functions of time.
The rotating neutron star on the stable branch remains stable throughout the simulation with oscillation amplitude $\sim 0.5\%$ of the central rest-mass density $\rho_c$, which agrees with the turning point theorem \cite{frie88}.
In contrast, the central rest-mass density $\rho_c$ of the unstable model (green solid lines) quickly drops and oscillates around the value of its stable counterpart, eventually damped and settling down to the stable state.
Despite the central rest-mass density of the unstable neutron star being only $6.6\%$ larger than the stable model,
the code can still resolve the migration of the unstable model remarkably well.
Since the oscillation is comparably small, no matter is ejected essentially during the migration.
As a result, the baryon mass $M_b$ and angular momentum $J$ are well conserved with a relative difference $\lesssim 10^{-12}$.

Nonetheless, if an initial perturbation is introduced in the unstable model (blue lines in \cref{fig:rns_migration_evo}),
the rotating neutron star immediately undergoes gravitational collapse due to the perturbation with an increasing central rest-mass density $\rho_c$.
After a short time, the neutron star compactness becomes so high that ultimately, a black hole is formed at $t_{\rm AH} = 0.931~{\rm ms}$ and swallows the whole star within the black hole,
leaving basically nothing outside the apparent horizon at the end, which agrees with the finding in \cite{shib03b}
(see the top right and bottom left panels in \cref{fig:rns_collapse_snapshot} for the profiles of the rest-mass density $\rho$ before and after the formation of the black hole).
The resultant black hole essentially inherits the initial neutron star's ADM mass and angular momentum with negligible loss.
The mass $M_{\rm BH}$ and angular momentum $J_{\rm BH}$ of the black hole extracted from the apparent horizon indeed show excellent agreement with derivation $\lesssim 0.03 \%$ as shown in \cref{fig:rns_migration_bh}.
This demonstrates the robustness and the accuracy of both the metric solver and the apparent horizon finder.
During the collapsing phase,
the baryon mass $M_b$ and angular momentum $J$ are conserved down to machine precision until the black hole is formed and the fluid excision is activated.
The Hamiltonian constraint violation $||\mathcal{H}||_2$ also experiences a sudden jump at $t_{\rm AH}$ due to the appearance of irregularity at the origin when the puncture is formed
and then quickly damped out and stabilized afterward.

\begin{figure}
    \centering
    \includegraphics[width=\linewidth]{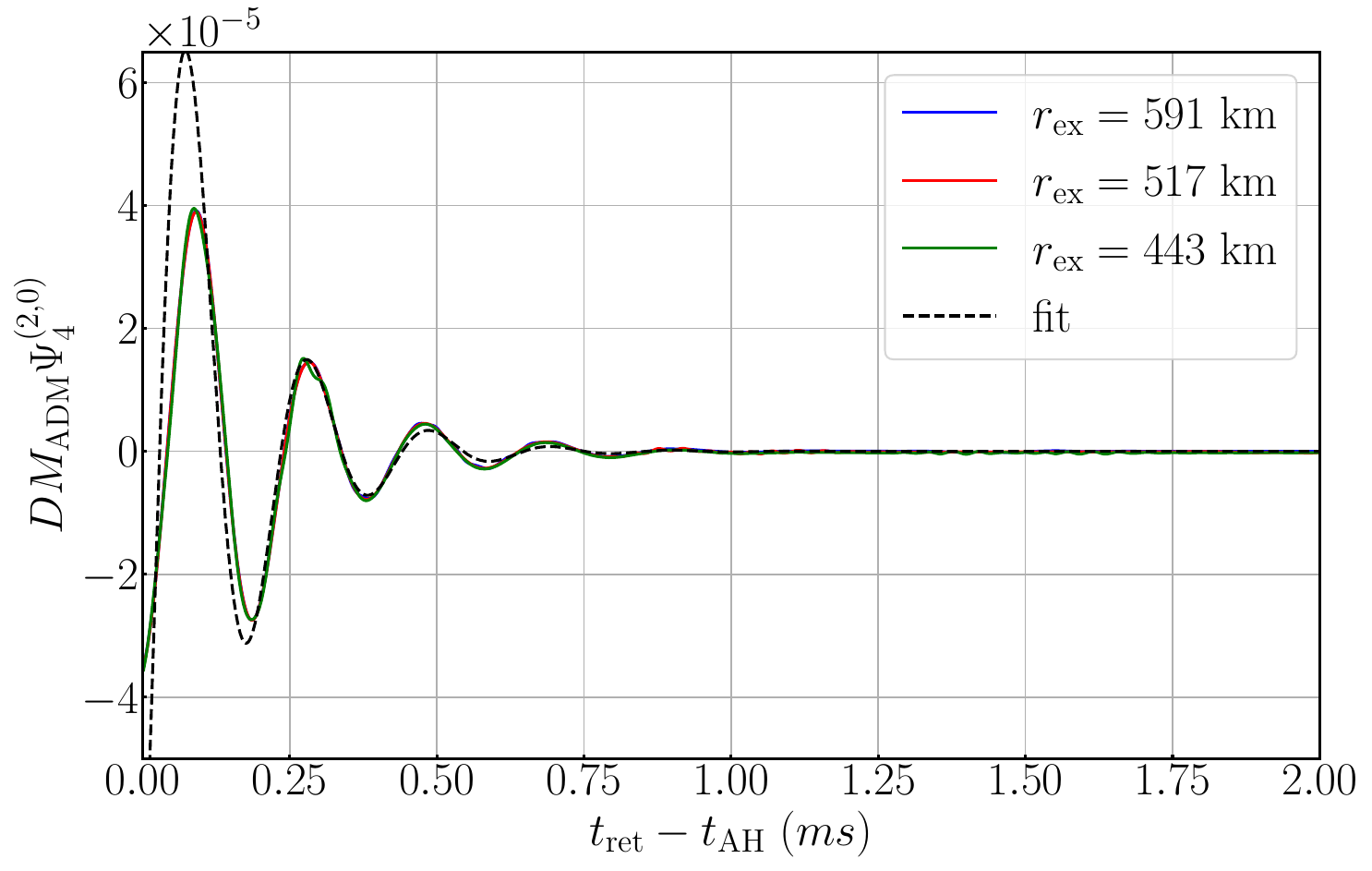}
    \caption{$(l,m)=(2,0)$ mode of $D M_{\rm ADM} \Psi_4$ gravitational waves extracted at $r_{\rm ex}= 591~{\rm km}$ (blue), $517~{\rm km}$ (red), and $443~{\rm km}$ (green) of the gravitational collapse of the unstable rotating neutron star as a function of retarded time $t_{\rm ret}$.
    }
    \label{fig:unstable_rns_psi4}
\end{figure}

In addition, we examine the gravitational wave signal from the collapse scenario of the perturbed unstable model.
Since the collapse happens promptly after the start of the simulation,
the initial junk radiation will contaminate the subsequent gravitational wave signal that immediately follows under the Z4c constraint propagating description.
Therefore, for this particular result shown in \cref{fig:unstable_rns_psi4}, we perform the simulation with the BSSN formulation to obtain a cleaner numerical waveform, and we confirm that the overall dynamics of the BSSN run are the same as in the Z4c run shown above.
The $(l,m)=(2,0)$ mode of $D\Psi_4$ black hole ringdown gravitational waves after the collapse is extracted in various radii as a function of $t_{\rm ret} - t_{\rm AH}$ shown in \cref{fig:unstable_rns_psi4},
where $t_{\rm ret}$ is the retarded time and $t_{\rm AH}$ is the black hole formation time.
The waveforms agree with each other regardless the extraction radii $r_{\rm ex}= 591~{\rm km}$, $517~{\rm km}$, and $443~{\rm km}$.
We also compare our numerical waveform with the analytical black hole quasinormal modes frequency 
$M_{\rm BH} \omega = 0.3767 - 0.0884i$ \cite{bert09}
considering the final black hole mass $M_{\rm BH}=2.51M_{\odot}$ and the dimensionless spin parameter $\chi = 0.2857$.
The fitted analytical ringdown waveform shown as the black dashed line in \cref{fig:unstable_rns_psi4} matches our result.
We found the total radiated energy to be $\sim 2.2 \times 10^{-9}M_{\rm ADM}$.

\subsubsection{Gravitational collapse of a supermassive star}
\begin{table}
\centering
\caption{The parameters of the SMS and the remnant black hole in the simulations.
$\Gamma$ is the adiabatic index,
$M$ is the gravitational mass of the system,
$\beta$ is the ratio of rotational kinetic energy to gravitational potential energy,
$J$ is the angular momentum
$R{\rm eq}$ is the equatorial circumferential radius, and
$M_{\rm BH}$ and $\chi$ are the mass and dimensionless spin of the remnant black hole, respectively.}
\begin{tabular}{
        >{\centering}p{0.13\columnwidth}
	>{\centering}p{0.17\columnwidth} 
        >{\centering}p{0.15\columnwidth} 
        >{\centering}p{0.10\columnwidth} 
        >{\centering}p{0.11\columnwidth} 
        >{\centering}p{0.14\columnwidth} 
        >{\centering\arraybackslash}p{0.10\columnwidth} 
    }
 \hline \hline \\[-.8em]
 $\Gamma$ & $M(M_\odot)$ & $\beta$ & $J/M^2$ & $R_{\rm eq} / M$ & $M_{\rm BH}/M$ & $\chi$ \\
 \hline \\[-.8em]
 1.3347 & $1.54\times 10^5$ & 0.00895 & 0.826 & 452.6 & 0.952 & 0.701 \\
 \hline \hline
\end{tabular}
\label{tab:sms}
\end{table}

For the final test, we simulate the gravitational collapse of a rotating supermassive star (SMS) to a black hole. In this problem, the SMS with a radius of $\approx 450M$ collapses to a black hole and a disk, and hence, we have to follow a much larger dynamical range than that of neutron-star collapses. For this problem, our FMR algorithm becomes, in particular, the robust tool. 

\begin{figure}
    \centering
    \includegraphics[width=\linewidth]{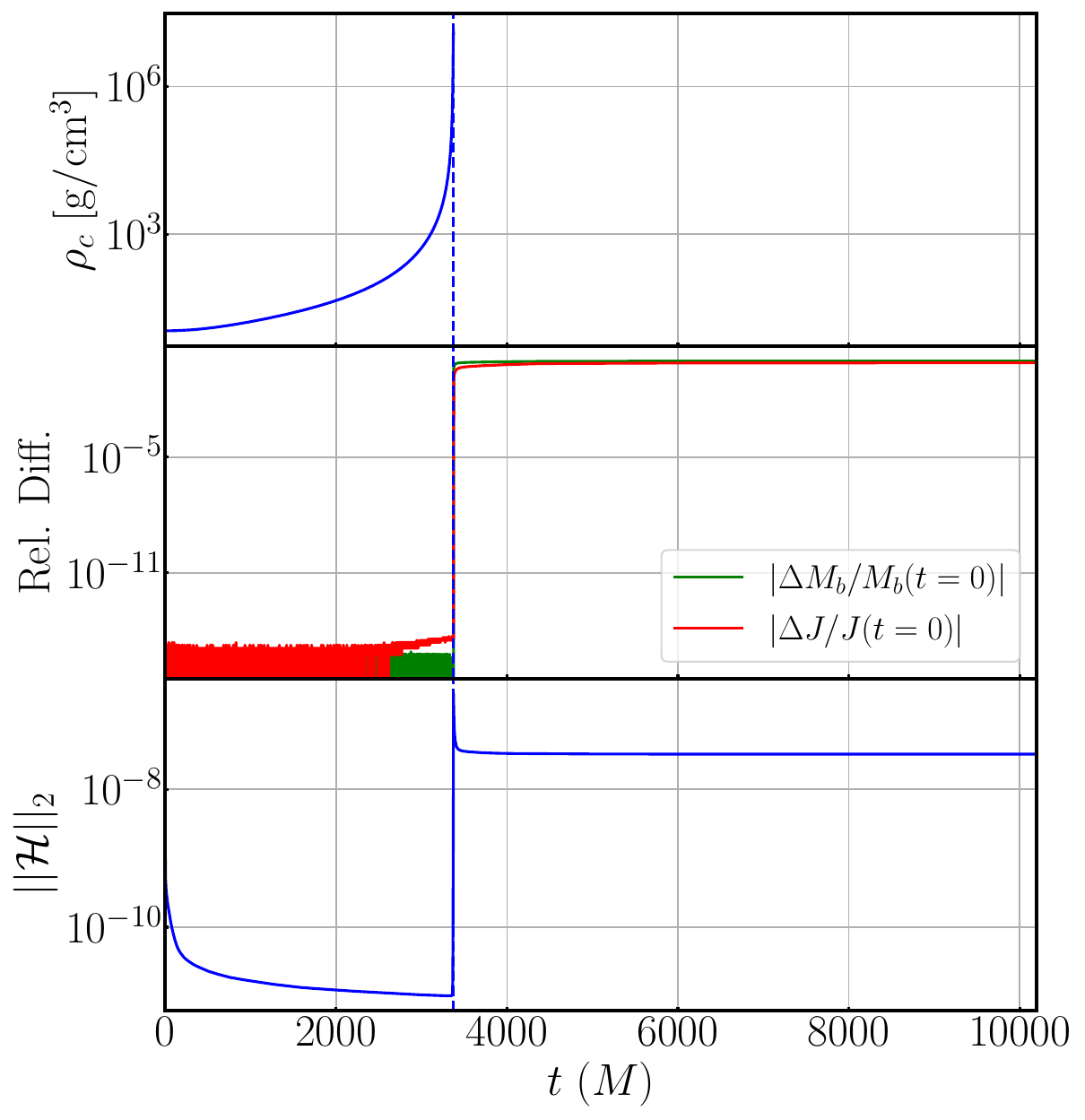}
    \caption{The evolution of the central rest-mass density $\rho_c$ (top), 
        relative differences in baryon mass $M_b$ and angular momentum $J$ (middle),
        and the L2-norm of the Hamiltonian constraint violation $||\mathcal{H}||_2$
        as a function of time $t$ in $M$.
        The blue vertical dashed line indicates the black hole formation time $t_{\rm AH} = 3373M$.}
    \label{fig:sms_evo}
\end{figure}
\begin{figure}
    \centering
    \includegraphics[width=\linewidth]{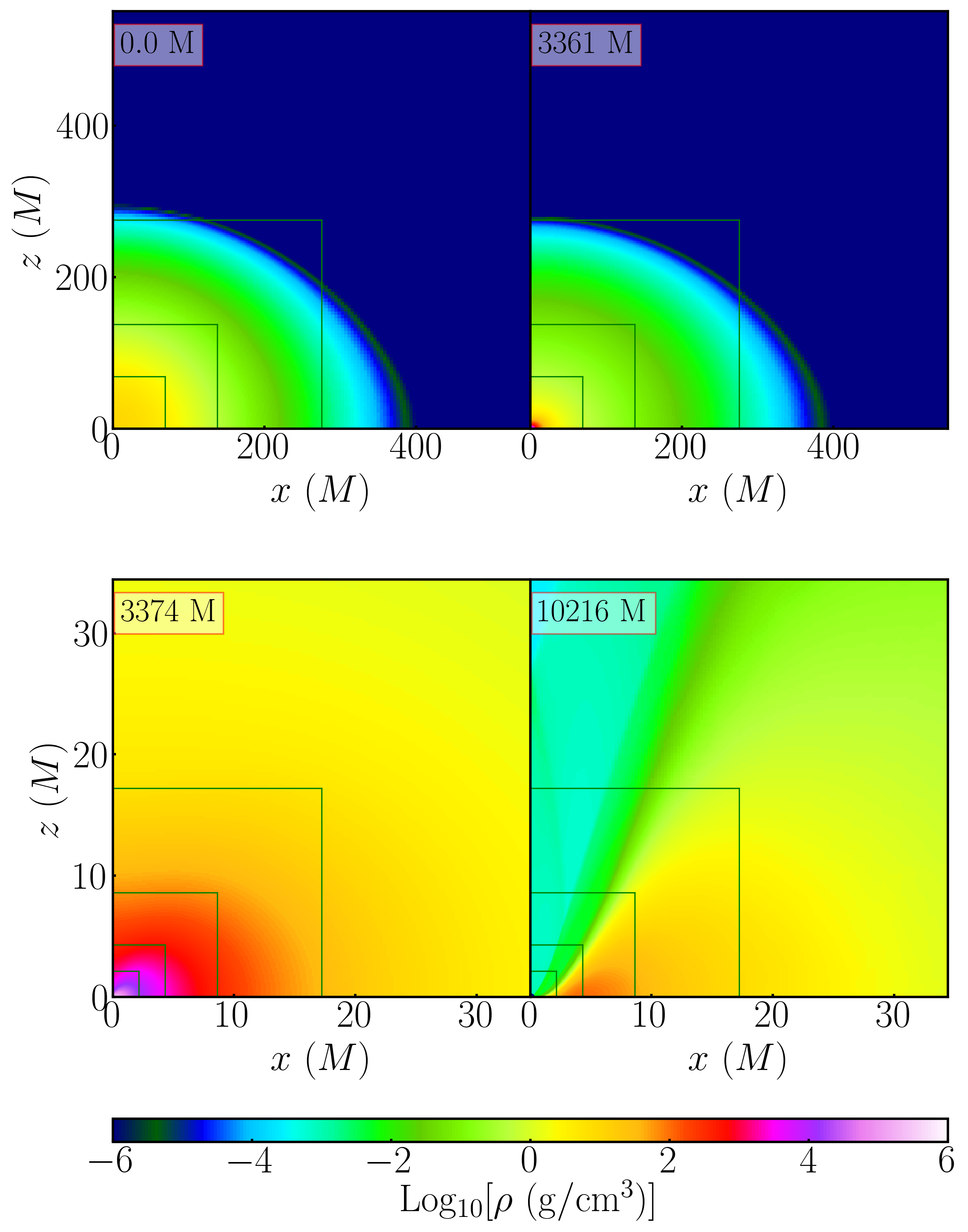}
    \caption{The snapshot of the rest-mass density $\rho$ extracted at different time
        $t = 0.0M$ (top left),
        $t = 3361M$ (top right),
        $t = 3374M$ (bottom left), and
        $t = 10216M$ (bottom right).
        The green solid lines indicate the boundaries of the FMR levels.}
    \label{fig:sms_snapshot}
\end{figure}
\begin{figure}
    \centering
    \includegraphics[width=\linewidth]{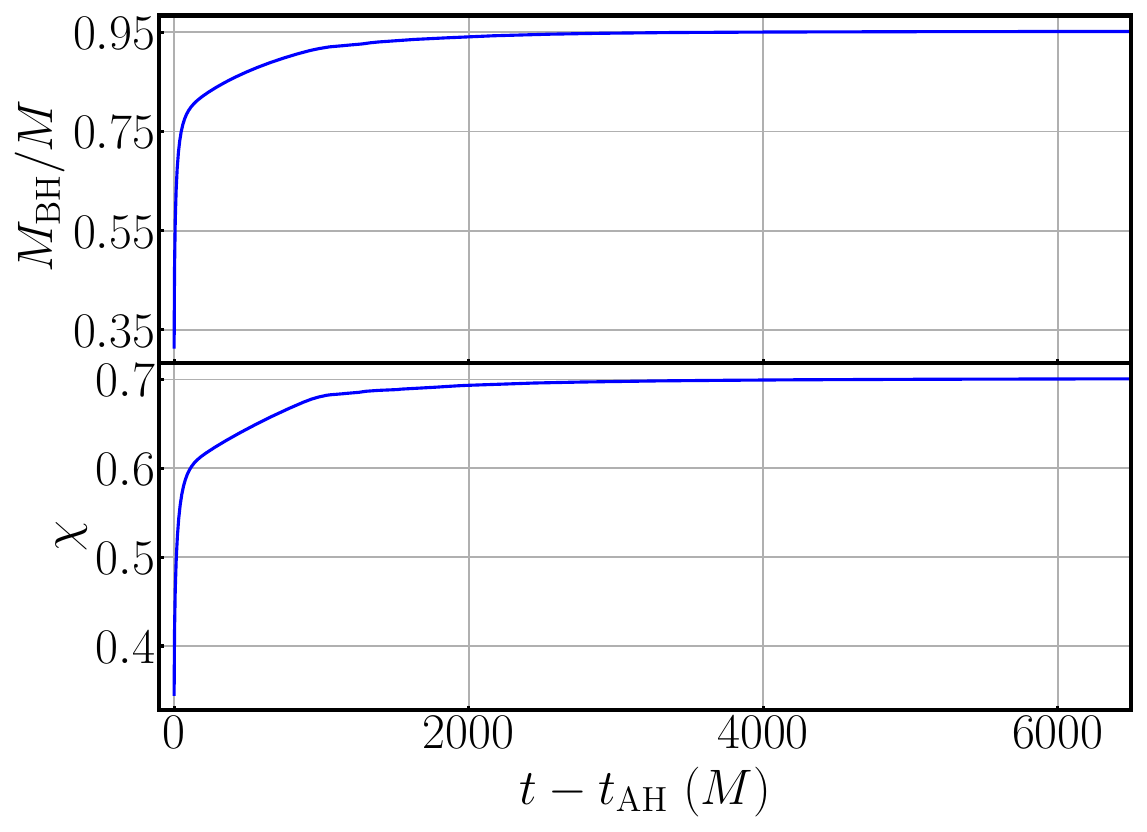}
    \caption{The evolution of remnant black hole mass $M_{\rm BH}$ (top) and dimensionless spin parameter $\chi$ (bottom) after collapse.
    The black hole is formed at $t_{\rm AH} = 3373M$.}
    \label{fig:sms_bh}
\end{figure}
\begin{figure}
    \centering
    \includegraphics[width=\linewidth]{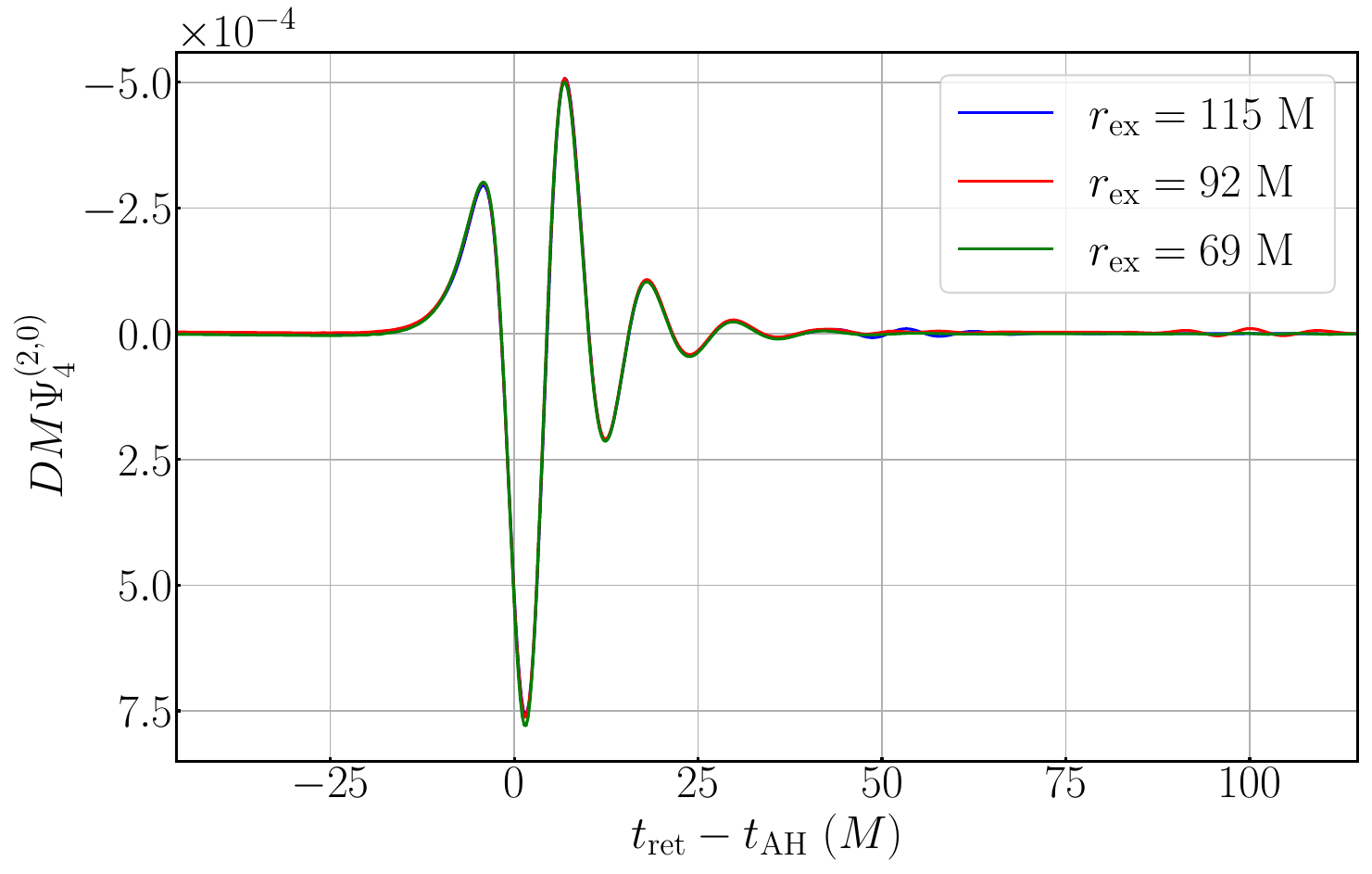}
    \caption{$(l,m)=(2,0)$ mode of $D M \Psi_4$ gravitational waves extracted at $r_{\rm ex}= 115M$ (blue), $92M$ (red), and $69M$ (green) of the gravitational collapse of the SMS as a function of retarded time $t_{\rm ret}-t_{\rm AH}$.
    }
    \label{fig:sms_psi4_2}
\end{figure}

We consider a uniformly rotating supermassive star constructed by the polytropic EOS $P=K \rho^\Gamma$ with the polytropic index $\Gamma = 1.3347$, which approximates the SMS core in helium-burning phase close to the marginally stable state~\cite{Shibata:2024xsl} and is approximately the same as the model He4 of \cite{fuji24b}.
The parameters of the model employed are listed in \cref{tab:sms}.

The computational domain is set to be $x_{\rm max}=z_{\rm max}=1101M$ with $N=128$ and 10 refinement levels in total,
which corresponds to the size of $L=2.15M$ and the grid resolution $\Delta x = \Delta z = 0.0168M$ in the finest box.
The $\Gamma$ law EOS $P = \rho (\Gamma - 1) \epsilon$ is employed for the simulation with the HLLC solve and the atmospheric factor $f_{\rm atm} = 10^{-20}$.
To initiate the collapse, we reduce the pressure by $20\%$ uniformly within the star. 
With our FMR setup, the computational cost for this simulation is relatively cheap, with the simulation time $t\approx 10800M$ costing about 600 CPUhrs in total under the parallelization setting $[M_{\rm MPI} \times M_{\rm MPI} \times N_{\rm thr} = [4 \times 4 \times 5]$.

Once the pressure is depleted, the matter starts to fall in, resulting in an exponential growth in the central rest-mass density $\rho_c$.
As in \cite{fuji24b}, about 95\% of the SMS collapses into a black hole, and the remaining matter forms a torus around the black hole and ejecta, which is driven by a shock formed around the surface of the torus as shown in the bottom row of the snapshots in \cref{fig:sms_snapshot}. The final dimensionless spin of the black hole is $\approx 0.70$, which is appreciably smaller than the dimensionless spin of the system (see \cref{tab:sms} and also \cref{fig:sms_bh} for the evolution of remnant black hole).
The ejecta mass is $\sim 1\%$ of the total mass, and this result agrees with that of \cite{fuji24b}.

Figure~\ref{fig:sms_psi4_2} plots gravitational waveform (the $(l,m)=(2,0)$ mode of $\Psi_4$) during the formation of the black hole. As found in \cite{shib16b}, the waveform is composed of a precursor, which is emitted before the formation of the black hole, a burst wave, which is emitted near the formation time of the black hole, and a ring down. The total radiated energy is $\approx 1.6 \times 10^{-6}M$. The order of the magnitude agrees with the result in \cite{shib16b}.

\subsection{Strong scaling test}\label{sec:scaling}

\begin{figure}
    \centering
    \includegraphics[width=\linewidth]{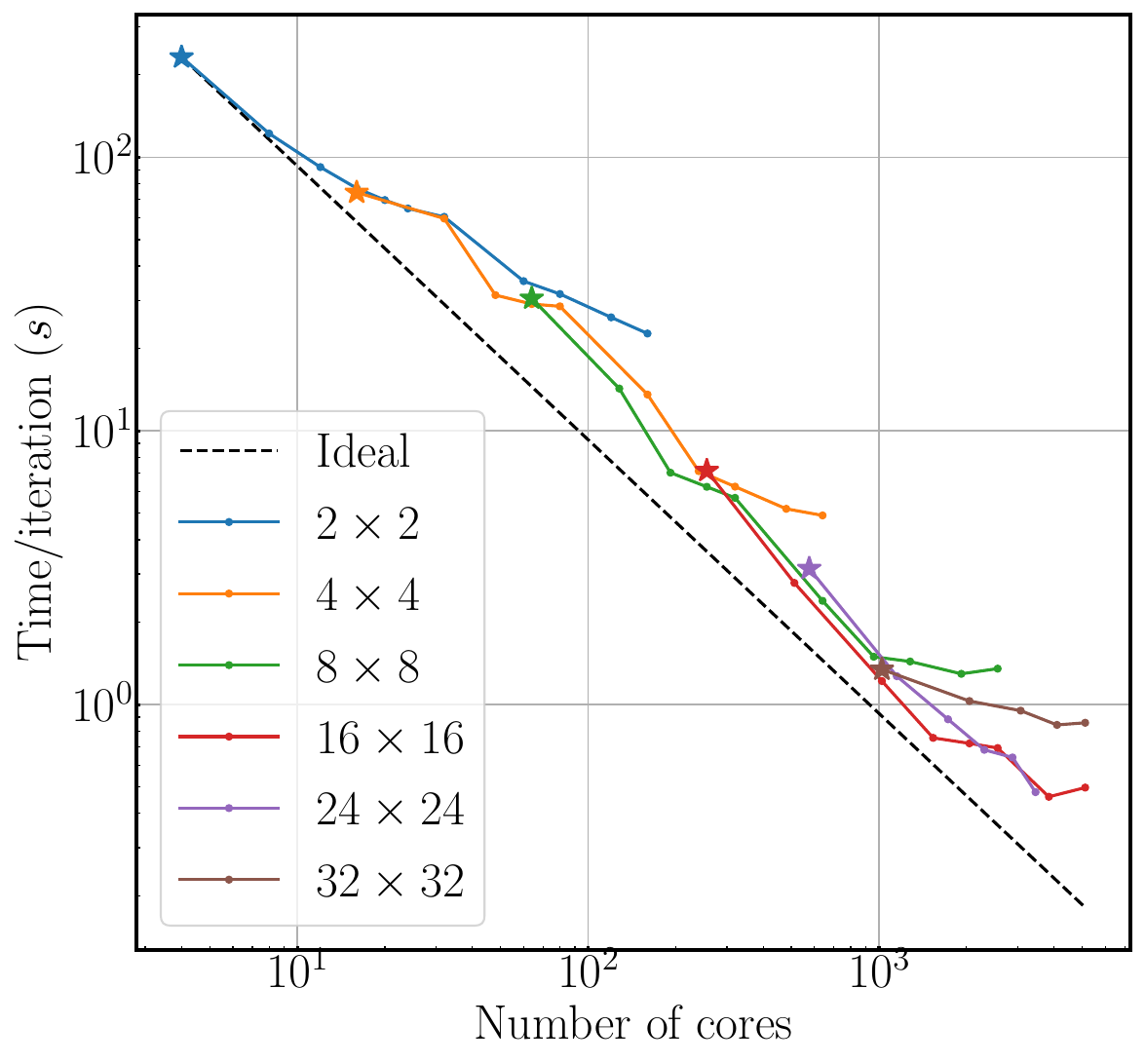}
    \caption{
        The result of the strong scaling test.
        The legend represents the MPI setting $[M_{\rm MPI}\times M_{\rm MPI}]$ for the models.
        The star markers indicate the result using a single OpenMP thread $N_{\rm thr}=1$, while we employed $N_{\rm thr} > 1$ for the dot markers.
    }
    \label{fig:scale}
\end{figure}

This section presents a test to assess the strong scaling of \texttt{SACRA-2D}.
The simulations were performed on the cluster Sakura at the Max Planck Computing and Data Facility,
which comprises Intel(R) Xeon(R) Gold 6248 CPU with a clock rate of 2.50\,GHz.
We performed a series of simulations in different parallelization settings using the same configuration in \cref{sec:stable_ns} except $N=960$ and 9 FMR levels were adopted here,
which corresponds to 35 cycles of RK4 integration in each time iteration,
and measured the average computational time required per iteration.
A wide range of MPI setting with the number of MPI ranks in each direction $M_{\rm MPI} \in \{ 2, 4, 8, 16, 24, 32 \}$
(see \cref{sec:parallel} for definition of $M_{\rm MPI}$ and $N_{\rm thr}$), covering 4 to 5120 cores in total.
\cref{fig:scale} shows the average computational time per iteration in seconds as a function of the number of cores used.
The solid line with the same colors denotes models with the same MPI setting but in different numbers of OpenMP threads $N_{\rm thr}$,
and the star markers represent the models with $N_{\rm thr} = 1$.
The black dashed line indicates the ideal scaling considering $2\times 2$ MPI setting with a single OpenMP thread $N_{\rm thr}=1$ (i.e. $4$ cores in total).
The result shows an efficiency of about $70\%$ for a small number of $N_{\rm thr}$,
and the performance worsens for an excessive number of OpenMP threads.
This suggests the optimal setting to be $N/M_{\rm MPI} \gtrsim 30$ for the MPI setting and $N_{\rm thr} \lesssim N/M_{\rm MPI}/10$ for the OpenMP threads.

\section{Summary}
We present \texttt{SACRA-2D}, a new MPI and OpenMP parallelized, fully relativistic hydrodynamics code in dynamical spacetime under axial symmetry with the cartoon method.
The code employs a cell-centered grid with FMR and an adaptive time-step scheme.
We implement the finite volume method with the state-of-the-art HLLC approximate Riemann solver for hydrodynamics
and the Baumgarte-Shapiro-Shibata-Nakamura formalism with Z4c constraint transport for spacetime evolution.

We examined \texttt{SACRA-2D} with several benchmark tests, including problems in the vacuum spacetime or the Cowling approximation and simulations of GRHD under dynamics spacetime.
We showed a sixth-order convergence of the metric solver and the gravitational waveform in the trumpet black hole and head-on collision tests, respectively.
We also demonstrated the power of the HLLC Riemann solver, which effectively improves spatial resolution in the modified Bondi flow test and reduces the artificial shock heating at the stellar surface in the simulation of a stable rotating neutron star. 
In particular, we show the outstanding robustness and efficiency of \texttt{SACRA-2D} in problems like examining the stability of rotating neutron stars adjoining the turning point and resolving the supermassive star collapse.
In addition, we performed a strong scaling test and showed an efficiency of about $70 \%$.

In the future, we plan to implement magnetohydrodynamics with the HLLD Riemann solver and the constrained transport scheme \cite{kiuc22}, as well as implementing radiation hydrodynamics for neutrino physics.
We will also use \texttt{SACRA-2D} to explore systems in the alternative theories of gravity.

\section*{Acknowledgement}
We thank the members of the Computational Relativistic Astrophysics department of the Max Planck Institute for Gravitational Physics for their helpful discussions.
The authors are indebted to Hao-Jui Kuan for preparing some initial data for the numerical tests and Kenta Kiuchi for revising this paper.
Numerical computation was performed on the cluster Sakura at the Max Planck Computing and Data Facility. This work was in part supported by Grant-in-Aid for Scientific Research (grant Nos.~20H00158 and 23H04900) of Japanese MEXT/JSPS.

\bibliographystyle{apsrev4-2}
\bibliography{references}

\end{document}